\documentclass[aps,preprintnumbers,floats, prd, twocolumn, nofootinbib]{revtex4-1}
\usepackage{graphicx}% Include figure files
\usepackage{dcolumn}% Align table columns on decimal point
\usepackage{bm}% bold math
\usepackage{amsmath}
\usepackage{hyperref}
\usepackage{comment}
\usepackage[utf8]{inputenc}
\usepackage{color}
\usepackage{orcidlink}
\usepackage{subfigure}
\hypersetup{colorlinks=true, citecolor=blue, linkcolor=blue, urlcolor=black}
\newcommand{\be}{\begin{equation}}
\newcommand{\ee}{\end{equation}}
\newcommand{\bea}{\begin{eqnarray}}
\newcommand{\eea}{\end{eqnarray}}

\DeclareRobustCommand{\Sec}[1]{Sec.~\ref{#1}}

\DeclareRobustCommand{\Fig}[1]{Fig.~\ref{#1}}
\DeclareRobustCommand{\Eq}[1]{Eq.~(\ref{#1})}

\DeclareRobustCommand{\r}[1]{{\rm #1}}

\newcommand{\tk}{T_\r{k}}
\newcommand{\xe}{x_\r{e}}
\newcommand{\tb}{T_{21}}
\newcommand{\sv}{\left<\sigma v\right>}
\newcommand{\rd}{\r{d}}
\newcommand{\mdm}{m_{\chi}}
\newcommand{\OmC}{\Omega_{\r{c}}}

\newcommand{\OmM}{\Omega_{\r{m}}}

\begin{document}

\title{Signatures of inhomogeneous dark matter annihilation on 21-cm}

\author{Junsong Cang\orcidlink{0000-0002-0061-0728}$^{1}$}
\email{cangjunsong@outlook.com}
\author{Yu Gao\orcidlink{0000-0002-8228-9981}$^{2}$}
\email{gaoyu@ihep.ac.cn}
\author{Yin-Zhe Ma\orcidlink{0000-0001-8108-0986}$^{4,5}$}
\email{mayinzhe@sun.ac.za}

\affiliation{$^1$ School of Physics, Henan Normal University, Xinxiang, China}
\affiliation{$^2$ Key Laboratory of Particle Astrophysics, Institute of High Energy Physics, Chinese Academy of Sciences, Beijing, 100049, China}
\affiliation{$^4$Department of Physics, Stellenbosch University, Matieland 7602, South Africa}
\affiliation{$^5$ National Institute for Theoretical and Computational Sciences (NITheCS), Stellenbosch University, Matieland 7602, South Africa}

\begin{abstract}

The energy released from dark matter (DM) annihilation leads to additional ionization and heating of the intergalactic gas, impacting the hydrogen 21-cm signal during the cosmic dawn. The dark matter annihilation rate scales with its density squared and becomes inhomogeneously boosted with structure formation. This paper examines the inhomogeneity in DM annihilation rate induced by the growth of DM halo structures, and we show that this effect can significantly amplify the spatial fluctuations in temperature and ionization fraction of the gas. Consequently, the fluctuations in the 21-cm brightness temperature may also be enhanced. We showcase these effects for a DM mass of 100 MeV annihilating into $\r{e}^-\r{e}^+$ at a rate of $\left<\sigma v\right>/m_\chi \sim 10^{-27} {\rm cm^3 s^{-1} GeV^{-1}}$, which is consistent with current constraints set by the cosmic microwave background. We find that, compared to the homogeneous calculations, inhomogeneous annihilation can enhance the 21-cm power spectrum by up to a factor of 130 over the scales of $k \in [0.05, 3]\ {\rm{Mpc^{-1}}}$ at redshifts $11-16$. Such signatures could potentially be detected by upcoming radio observatories such as the Square Kilometer Array telescope.

\end{abstract}

\maketitle

\section{Introduction}

Dark matter (DM) annihilation events can produce secondary particles that can potentially be detected through astrophysical probes~\cite{Slatyer:2021qgc,Hooper:2018kfv}.
During their propagation in the Universe,
secondary particle cascades from DM annihilation result in additional ionization and heating of the intergalactic medium (IGM).
Such ionization effects enhance the scattering between the cosmic microwave background (CMB) photons and free electrons, 
leading to observable effects in the CMB anisotropy spectrum~\cite{Planck:2018vyg,Padmanabhan:2005es,Slatyer:2012yq,Slatyer:2015jla,Slatyer:2015kla,Liu:2016cnk}.
The latest measurements by {\it Planck}~\cite{Planck:2018vyg} have placed DM annihilation constraints that are competitive with those from high-energy cosmic-ray searches~\cite{Slatyer:2021qgc,Hooper:2018kfv,MAGIC:2016xys,Bergstrom:2013jra,Giesen:2015ufa}.
The heating effect from DM annihilation, on the other hand,
can be efficiently probed by future 21-cm observations on cosmic neutral hydrogen.

The 21-cm signal arises from the transition between the singlet and triplet states of neutral hydrogen, 
it offers an invaluable glimpse into the epochs from Cosmic Dawn to Reionization~\cite{Pritchard:2011xb}.
During these epochs, the formation of DM halo structures is expected to enhance the DM annihilation rate significantly.
The 21-cm signal strength is sensitive to the thermal and ionization conditions in the IGM,
thereby providing a unique avenue to detect possible heating and ionization signatures from DM annihilation.
The primary 21-cm observation window below redshift $20$ lies deep in the epoch of nonlinear structure growth, during which spatial inhomogeneity is expected in the DM halo-boosted heating and ionizing sources. This leads to modifications in the 21-cm power spectrum at scales where spatial inhomogeneity manifests itself.

In this paper, 
we examine the effects of DM distribution inhomogeneity on the 21-cm power spectrum.
We will show that the formation of DM halo structures induces inhomogeneities in the DM annihilation rate that closely trace the density fluctuations.
Assuming that DM annihilation products have a short absorption length,
which is generally feasible for relatively light dark matter and the resultant radiation is at low energy,
we find that the DM-induced heating and ionization exhibit distinctive inhomogeneous structures,
which can further enhance spatial fluctuations and the power spectrum of 21-cm signal.
Such features can be particularly helpful in discriminating potential DM signatures from complex astrophysical backgrounds.
For an annihilation rate of $\sv /\mdm \sim 10^{-27} \r{cm^3 s^{-1} GeV^{-1}}$,
which is roughly the same level as that constrained by cosmic-rays and CMB~\cite{Planck:2018vyg,Slatyer:2021qgc,Hooper:2018kfv,MAGIC:2016xys},
the corresponding 21-cm power spectrum can be easily detected by the Square Kilometer Array (SKA) telescope~\cite{Sitwell:2013fpa}.

This paper is organized as follows:
Sec.~\ref{juhhdyf83_dhfdgvt} briefly reviews the basics of cosmic 21-cm signal,
in Sec.~\ref{Section_2} we describe our inhomogeneous DM annihilation model, Sec.~\ref{juhhdyf83_dhfdsgvt} present our results on the 21-cm power spectrum, and then we conclude in Sec.~\ref{juhhdsayf88hhfe3_dhfdsgvt}.

\section{Cosmic 21-cm signal}
\label{juhhdyf83_dhfdgvt}

The 21-cm signal arises from the hyperfine energy splitting between the singlet and triplet states of neutral hydrogen atoms.
In cosmological context,
the signal strength is measured by the 21-cm brightness temperature $T_{21}$~\cite{Mesinger:2010ne,Pritchard:2011xb},
\begin{eqnarray}
T_{21}
& \simeq &
27 x_\r{HI}
\left(
1 + \delta_{\rm b}
\right)
\left(
\frac{H}
{
\rd v_\r{r}
/
\rd r
+
H
}
\right)
\left(
1 - 
\frac{T_\gamma}{T_\r{s}}
\right) \nonumber
\\
&\times &
\left(
\frac{1+z}{10}
\frac{0.15}
{\Omega_\r{m}h^2}
\right)^{1/2}
\left(
\frac{\Omega_\r{b}h^2}
{0.023}
\right)
\r{mK},
\label{Eq_T21_definition}
\end{eqnarray}
where $z$ is the redshift,
$x_\r{HI}$ is the neutral fraction of the IGM,
$\delta_{\rm b}$ is the baryon density contrast,
$H(z)$ is the Hubble parameter,
$\rd v_\r{r}/\rd r$ is the velocity gradient along the line of sight and
$T_\gamma$ indicates the temperature of the radiation background,
which is typically assumed to be the CMB temperature.
$\Omega_{\rm m}$ and $\Omega_{\rm b}$ are the fractional densities for matter and baryon respectively (therefore fractional density of cold dark matter is $\Omega_{\rm c}=\Omega_{\rm m}-\Omega_{\rm b}$),
$h$ is the Hubble constant in unit of $100\ \r{km/s/Mpc}$.
The spin temperature $T_\r{s}$ quantifies the number ratio of hydrogen atoms occupying the singlet and triplet states, and it is coupled to both the radiation temperature $T_\gamma$ and the gas kinetic temperature $T_\r{k}$ through collisions and the Wouthuysen-Field effect~\cite{Furlanetto:2006jb,Pritchard:2011xb,Mesinger:2010ne},
\be
T^{-1}_\r{s}
=
\frac{
T^{-1}_\gamma
+
x_\alpha
T^{-1}_\alpha
+
x_\r{c}
T^{-1}_\r{k}
}
{1 + x_\alpha + x_\r{c}},
\label{dshfhiy3yedvdsf4r78r6tfgy}
\ee
here the color temperature $T_\alpha$ is typically closely coupled to $T_\r{k}$,
$x_\r{c}$ is collisional coupling coefficient, and $x_\alpha$ describes Wouthuysen-Field coupling~\cite{Mesinger:2010ne},
\be
x_\alpha = \frac{1.7 \times 10^{11}}{1+z}
\left(
\frac{J_\alpha}
{\r{s^{-1} Hz^{-1} cm^{-2} sr^{-1}}}
\right)S_\alpha,
\ee
where $S_\alpha$ is a quantum mechanical correction of the order of unity,
and $J_\alpha$ is the Lyman-Alpha flux,
\be
J_\alpha = J_{\alpha, \r{astro}} + J_{\alpha, \chi}
\ee
here the first and second terms on the right hand side denote contributions from astrophysical sources and dark matter ($\chi$) respectively.
Calculation of $J_{\alpha, \r{astro}}$ is detailed in~\cite{Mesinger:2010ne},
and we compute the DM term following~\cite{Lopez-Honorez:2016sur,Facchinetti:2023slb}:
\be
J_{\alpha, \chi}
=
\frac{1}{4 \pi H \nu_\alpha E_\alpha}
\left[
\frac{\rd E}
{\rd V \rd t}
\right]_\r{dep, LyA}
\label{Eq_J_LyA_DM}
\ee
where $H$ is the Hubble rate,
$\nu_\alpha$ is the Lyman-Alpha frequency,
$E_\alpha = 10.2$ eV is the energy of a Lyman-Alpha photon,
$\left[\rd E /\rd V \rd t \right]_\r{dep, LyA}$ is the energy deposited into the Lyman-Alpha (LyA) channel per unit volume and time,
we will discuss this term in detail in \Sec{Subsec_Recombination_Equations}.

As can be seen from \Eq{Eq_T21_definition},
spatial fluctuations in gas ionization,
density,
temperature,
and velocity gradient will make $T_{21}$ inherently inhomogeneous.
Therefore, 
in addition to the global average $\bar{T}_{21}$,
the 21-cm signal is also characterized by its spherically averaged power spectrum $P_\phi$~\cite{Lopez-Honorez:2016sur}:
\be
\left<
\tilde{\delta_\phi}
(\vec{k},z)
\tilde{\delta_\phi^{*}}
(\vec{k'},z)
\right>
\equiv
(2 \pi) ^3
\delta^{(3)}_\r{D}
\left(\vec{k} - \vec{k'} \right)
P_\phi(k,z),
\ee
where $\phi$ denotes the physical quantity under consideration,
which can take values of $T_{21}$,
density field $\delta$,
and boost factor $B$ (see next section) in the context of this work.
The brackets $\left< \right>$ denote the ensemble average,
$\tilde{\delta_\phi}({\vec{k}}, z)$ represents the Fourier transform of $\delta_\phi({\vec{x}}, z) \equiv \phi ({\vec{x}}, z)/\bar{\phi}(z) - 1$,
and $\bar{\phi}(z)$ is the spatially averaged value of $\phi ({\vec{x}}, z)$.
$\delta_\r{D}$ is the three-dimensional Dirac function, and
$k=|\vec{k}|$.
In practice,
hereafter we will use the reduced power spectrum $\bar{\phi}^2 \Delta^2_\phi$,
where $\Delta^2_\phi$ is defined through,
\be
\Delta^2_\phi
(k,z)
\equiv
\frac{k^3}
{2 \pi^2}
P_\phi(k,z),
\ee
We compute all power spectra in our analysis using the {\tt powerbox} package~\cite{2018JOSS....3..850M}.

Equations~(\ref{Eq_T21_definition}) and (\ref{dshfhiy3yedvdsf4r78r6tfgy}) show that the 21-cm signal is encoded with information about the thermal and ionization states of the IGM ($T_\r{k},\ x_\r{HI}$).
For the epochs of interest ($5 \lesssim z \lesssim 40$),
the IGM is affected by energy injection from annihilating DM as well as the radiation from the first galaxies,
therefore discrimination of possible DM signal using $T_{21}$ requires thorough knowledge of the astrophysical background.
Our calculations are built primarily on the {\tt 21cmFAST} code~\cite{Mesinger:2010ne,Park:2018ljd},
which is a fast semi-analytic package for simulating both the density field and astrophysical radiation,
interested readers are referred to~\cite{Park:2018ljd,Mesinger:2010ne} for review and program details.

Throughout this work,
we adopt a $\Lambda$CDM cosmology with {\it Planck} 2018 parameters~\cite{Planck:2018vyg}.
For the background astrophysics settings in {\tt 21cmFAST},
\footnote{Different {\tt 21cmFAST} versions have different implementations for astrophysical models,
here we use version 3.2.1,
which contains implementation of astrophysical model in~\cite{Park:2018ljd}.},
we use the same values as that in~\cite{Park:2018ljd},
which has been shown to be consistent with measurements of the UV luminosity function~\cite{Bouwens:2014fua,Bouwens:2015vha,Oesch_2018},
optical depth~\cite{Planck:2016mks} and reionization timing~\cite{McGreer:2014qwa}.

\section{Inhomogeneous dark matter annihilation}
\label{Section_2}

Assuming that DM particles $\chi$ annihilate through $s$-wave with a thermally averaged cross-section $\sv$,
the energy injected per unit volume and time (referred to as injection rate hereafter for convenience) can be written as,
\be
\left[
\frac{\rd E}
{\rd V \rd t}
\right]
 =
2 \mdm
\cdot
g
\sv
n^2_{\chi}
=
\frac{\sv}
{m_{\chi}}
\rho^2_{\r{c}},
\label{ewd9gyfsd3}
\ee
where $\mdm$ denotes the DM mass,
$\rho_\r{c}$ is the DM density,
$n_\chi = \rho_\r{c}/m_\chi$ is the number density of DM particle,
and $\sv n^2_{\chi}$ is the number of DM annihilation events per unit volume and time,
$g$ is a symmetry factor which we take $1/2$ following~\cite{Planck:2018vyg}.
For homogeneous distribution,
using $\rho_\r{c} = \bar{\rho}_\r{c} = \OmC \rho_\r{cr} (1+z)^3$, where $\rho_\r{cr}$ is the current critical density,
\Eq{ewd9gyfsd3} can be expressed as,
\be
\left[
\frac{\rd E}
{\rd V \rd t}
\right]_{\r{HMG}}
=
\frac{\sv}{\mdm}
\Omega^2_{\rm{c}}
\rho^2_{\r{cr}}
(1+z)^6,
\label{fdgjrfu498rgfn}
\ee
where the subscript HMG denotes homogeneous distribution.

As can be seen from \Eq{ewd9gyfsd3},
the global injection rate of DM is proportional to $\bar{\rho^2}_\r{c}$,
which is simply $\bar{\rho_\r{c}}^2 = \Omega^2_{\rm{c}} \rho^2_{\r{cr}}(1+z)^6 $ for homogeneous distribution.
As matter overdensities grow at lower redshifts ($z \leq 50$),
the assumption of homogeneity in \Eq{fdgjrfu498rgfn} is no longer valid,
and $\bar{\rho^2_\r{c}}$ can exceed $\bar{\rho_\r{c}}^2$ by orders of magnitude.
As a result, 
the overall DM injection rate can also be significantly enhanced.
Therefore at low redshifts we model DM injection by combining  the contributions from both collapsed halos and from un-collapsed regions.

\subsection{Collapsed Halos}
The total DM annihilation rate per unit volume inside collapsed halos can be obtained by summing the contributions from individual halos.
The bolometric luminosity $L_\r{DM}$ from DM annihilating inside a halo can be calculated by integrating \Eq{ewd9gyfsd3} within the halo volume,
\be
L_\r{DM}
=
\frac{4 \pi \sv}{\mdm}
\int
\rd r r^2 \rho^2_{\r{c,halo}}(r),
\label{dsfefe78gfr4}
\ee
where $r$ is the distance to the halo center.
$\rho_\r{c,halo}(r)$ is radial DM density profile of the halo,
for which we adopt the Navarro-Frenk-White (NFW) parametrization~\cite{Navarro:1996gj,Ziparo:2022fnc, Maccio:2006wpz},
\be
\rho_\r{c,halo}(r)
=
\frac{\rho_\r{cr} \delta_\r{c} r_\r{vir}}
{c r (1+cr/r_\r{vir})^2}
\left(\frac{H}{H_0}\right)^2
,
\ee
here $H_0$ is the Hubble constant,
$G$ is the gravitational constant, 
and $r_\r{vir}$ is the virial radius~\cite{Ziparo:2022fnc,Barkana:2000fd}
\begin{eqnarray}
r_\r{vir}
&= &
0.784
\left(
\frac{m}
{10^8 h^{-1} m_\odot}
\right)^{1/3}
\left(
\frac{\OmM}
{\OmM^z}
\frac{\Delta_\r{c}}
{18 \pi^2}
\right)^{-1/3}
\nonumber \\
&\times &
\left(
\frac{10}{1+z}
\right)
h^{-1}
\r{kpc},
\end{eqnarray}
where $\Delta_\r{c}$ is the mean overdensity of the halo relative to mean density and is given by
$\Delta_\r{c} = 18 \pi^2 + 82 d - 39 d^2$,
$d = \OmM^z - 1$ and $\OmM^z = \OmM (1+z)^3/(\OmM (1+z)^3 + \Omega_\Lambda)$,
$\Omega_\Lambda$ is the fractional density parameter for dark energy.
$\delta_\r{c}$ is related to $\Delta_\r{c}$ as~\cite{Navarro:1996gj},
\be
\delta_\r{c}
=
\frac{\Delta_\r{c}}
{3}
\frac{c^3}
{
\ln(1+c)
-
c/(1+c)
},
\ee

Following the results derived from N-body simulation in~\cite{Sanchez-Conde:2013yxa},
we compute the halo concentration parameter $c$ as,
\be
c
=
\frac{1}{1+z}
\sum_{i=0}^5
c_i
\left[
\ln
\left(
\frac{m}
{h^{-1} m_\odot}
\right)
\right]^i,
\label{Eq_concentration}
\ee
where $c_i = [37.5153, -1.5093, 1.636\cdot 10^{-2}, 3.66 \cdot 10^{-4}, -2.89237\cdot10^{-5}, 5.32 \cdot 10^{-7}]$ are fitting coefficients.
Compared with the power-law extrapolation using the results derived from simulations of massive halos ($10^{10} \lesssim m/m_\odot \lesssim 10^{15}$, see e.g., \cite{Maccio:2006wpz,Maccio:2008pcd}),
which can lead to significant over-estimate of boost factor~\cite{Sanchez-Conde:2013yxa},
\Eq{Eq_concentration} provides accurate description for concentration parameter in a broad mass range of $[10^{-6}, 10^{15}]h^{-1} m_\odot$~\cite{Sanchez-Conde:2013yxa}.

For each simulation cell,
the injection rate from DM annihilating in halos is given by,
\be
\begin{aligned}
\left[
\frac{\rd E}
{\rd V \rd t}
\right]_{\r{Halo}}
&=
(1+z)^3
\int_{m_\r{min}} \rd m
\frac{\rd n}{\rd m}(\delta)
\cdot
L_\r{DM}
\\
&=
\frac{4 \pi \sv(1+z)^3}{\mdm}
\\
&
\times
\int_{m_\r{min}} \rd m
\frac{\rd n}{\rd m}(\delta)
\left[
\int
\rd r r^2 \rho^2_{\r{c,halo}}
\right],
\end{aligned}
\label{Eq_dEdVdt_Halo}
\ee
while deriving the second line we have used \Eq{dsfefe78gfr4},
the prefactor of $(1+z)^3$ converts injection rate from comoving frame to physical frame,
$\delta(x)$ is the density contrast at the cell's location,
$m_\r{min}$ is the minimum mass below which free-streaming prevents the formation of DM halos.
$m_\r{min}$ is dependent on DM properties,
namely the mass, 
cross-sections for annihilation,
scattering with baryons and self-scattering and can range between $10^{-10}$ to $10^2m_\odot$~\cite{Liu:2016cnk,Bringmann:2009vf,Profumo:2006bv,vandenAarssen:2012ag,Hofmann:2001bi}.
Following Ref.~\cite{Liu:2016cnk},
we adopt $m_\r{min} = 10^{-6}m_\odot$,
which is the canonical value for WIMP (Weakly Interacting Massive Particles) dark matter.

Finally in \Eq{Eq_dEdVdt_Halo},
$\rd n / \rd m (\delta)$ represents the conditional halo mass function (HMF),
which describes the comoving dark matter halo number density per unit mass interval for a region with overdensity $\delta$.
Following the {\tt 21cmFAST} treatments for astrophysical radiation fields~\cite{Mesinger:2010ne,Park:2018ljd},
we calculate $\rd n / \rd m (\delta)$ by normalizing the conditional Press-Schechter HMF~\cite{Lacey:1993iv,Somerville:1997df,Cooray:2002dia} computed with {\tt 21cmFAST} from excursion set theory to match the mean of the Sheth–Tormen HMF~\cite{Sheth:2001dp,Sheth:1999mn}
\footnote{
With this normalization our boost factor (defined in \Eq{Eq_Boost_factor_definition}) takes the form,
\be
B(x)
=
B_\r{ST}
\frac{B_\r{PS}(x)}
{\bar{B}_\r{PS}},
\ee
where $B_\r{ST}$ and $B_\r{PS}(x)$ represent the boost factors computed with Sheth–Tormen HMF and Press-Schechter HMF respectively,
$\bar{B}_\r{PS}$ is the spatial average of $B_\r{PS}(x)$.
},
and in computing the matter power spectrum we use the transfer function in~\cite{efstathiou1992cobe}.

\subsection{Uncollapse IGM regions}
Outside the collapsed halos,
we compute the DM density $\rho_\r{c}$ as,
\be
\rho_\r{c}(x)
=
(1 - f_\r{coll})
\bar{\rho}_\r{c},
\label{fdg543hgjhfddfegrhreh}
\ee
where $f_\r{coll}$ is the fraction of matter collapsed into halos and can be computed from conditional halo mass function as,
\be
f_\r{coll}
(x)
=
\frac{1}{\Omega_\r{m} \rho_{\r{cr}}(1+\delta)}
\int_{m_\r{min}} \rd m
\cdot
m\frac{\rd n}{\rd m}
(\delta).
\ee

Inserting \Eq{fdg543hgjhfddfegrhreh} into \Eq{ewd9gyfsd3},
we obtain the injection rate from DM annihilating in uncollapsed IGM regions,
\be
\left[
\frac{\rd E}
{\rd V \rd t}
\right]_\r{IGM}
=
\frac{\sv}{\mdm}
\left[1-f_\r{coll}(x)\right]^2
\Omega^2_{\rm{c}}
\rho^2_{\r{cr}}
(1+z)^6.
\label{dfgy4iurfy4ghcdg}
\ee

\subsection{Inhomogeneous Boost Factor}
At low redshifts,
the net DM injection rate is the sum of DM annihilating in halos and in the uncollapsed IGM,
and we parameterize the enhancement relative to injection in the uniform frame using the boost factor $B(x)$ as defined below,
\be
B(x)
\equiv
\left(
\left[
\frac{\rd E}{\rd V \rd t}
\right]_{\r{Halo}}
+
\left[
\frac{\rd E}{\rd V \rd t}
\right]_{\r{IGM}}
\right)
\Big/
\left[
\frac{\rd E}{\rd V \rd t}
\right]_{\r{HMG}}
\label{Eq_Boost_factor_definition}
\ee
from which the accurate injection rate can be recovered as,
\be
\left[
\frac{\rd E}{\rd V \rd t}
\right](x)
=
B(x)
\left[
\frac{\rd E}{\rd V \rd t}
\right]_{\r{HMG}}.
\ee
Using Eqs.~(\ref{fdgjrfu498rgfn},\ref{Eq_dEdVdt_Halo}) and \Eq{dfgy4iurfy4ghcdg}
\footnote{
For the NFW profile,
the integration for halo density profile can be solved analytically as
\begin{eqnarray}
\int {\rm d} r r^2 \rho^2_{\r{c,halo}}
= \frac{\rho^{2}_{\rm cr}\delta^{2}_{\rm c}r^{3}_{\rm vir}H^2}{3c^{3}(1+c)^{3}H^2_0}
\left[(1+c)^{3}-1 \right].
\end{eqnarray}
},
\be
\begin{aligned}
B(x)
=&
\left[1-f_{\r{coll}}(x)\right]^2
+
\frac{4 \pi}{\Omega^2_{\r{c}} \rho^2_{\r{cr}} (1+z)^3}
\\
&
\times
\int_{m_\r{min}}
\rd m
\frac{\rd n}{\rd m}(\delta)
\left[
\int \rd r r^2 \rho^2_{\r{c,halo}}
\right].
\end{aligned}
\label{mmbhxshdu3uutdjshyf1eierhdcdc}
\ee

Note that the boost factor in the cosmological context has been studied in literatures 
(see~\cite{Taylor:2002zd,Huetsi:2009ex}, etc.),
and here in \Eq{mmbhxshdu3uutdjshyf1eierhdcdc} we use the conditional halo mass function,
which encodes information about density fluctuations and therefore allows us to derive the inhomogeneous DM injection rate and its impact on 21-cm power spectrum.
Our comprehensive treatment of the {\it inhomogeneous boost factor} represents the main modeling improvement in this work.

\subsection{Recombination equations}
\label{Subsec_Recombination_Equations}

The energy injected from annihilating DM can in general be absorbed by the IGM and cause additional ionization and heating. 
The evolution equations for ionization fraction $x_\r{e}$ and gas kinetic temperature $T_\r{k}$ now become~\cite{Cang:2021owu,Liu:2016cnk},
\be
\begin{aligned}
\frac{\rd x_\r{e}}
{\rd t}
(x)
=&
\left[
\frac{\rd x_\r{e}}
{\rd t}
\right]_\r{Fiducial}
+
\frac{f_\r{ion}}
{(n_\r{H} + n_\r{He})E_\r{i}}
\left[
\frac{\rd E}{\rd V \rd t}
\right](x)\\
&+\left(1-C\right)
\frac{f_\r{LyA}}
{(n_\r{H} + n_\r{He}) E_{\alpha}}
\left[
\frac{\rd E}{\rd V \rd t}
\right](x),
\end{aligned}
\label{Eq_dxedt}
\ee

\be
\frac{\rd T_\r{k}}
{\rd t}
(x)
=
\left[
\frac{\rd T_\r{k}}
{\rd t}
\right]_\r{Fiducial}
+
\frac{2 f_\r{heat} \left[
{\rd E}/{\rd V \rd t}
\right](x)}
{3 k_\r{B}
n_\r{H}
(1+x_\r{e} + f_\r{He} + x_\r{e} f_\r{He})
},
\label{Eq_dTdt}
\ee
here the subscript ``Fiducial'' indicates the background evolution in absence of DM injection,
which has been detailed in Refs.~\cite{Park:2018ljd,Mesinger:2010ne}.
The ionization fraction is defined in {\tt 21cmFAST}\footnote{The ionization fraction definition used in {\tt 21cmFAST} is slightly different from that in typical recombination codes, e.g., {\tt HyRec}~\cite{Ali-Haimoud:2010hou,Lee:2020obi},
which uses convention $x_\r{e} \equiv n_\r{e} / n_\r{H}$.} as $x_\r{e} \equiv n_\r{e}/(n_\r{H} + n_\r{He})$~\cite{Mesinger:2010ne,Sarkar:2022dvl},
where $n_\r{e}$ and $n_\r{H}$ are number densities of free electrons and hydrogen nuclei respectively,
helium nuclei number density $n_\r{He}$ is given by $n_\r{He} = f_\r{He} n_\r{H}$,
and $f_\r{He} \equiv Y_\r{He}/[4(1-Y_\r{He})] \simeq 0.08$ is helium number fraction relative to hydrogen,
$Y_\r{He}$ is helium mass fraction for which we set $Y_\r{He} = 0.245$~\cite{Planck:2018vyg}.
$n_\r{H}$ is related to overdensity $\delta$ by $n_\r{H} = (1+\delta) \bar{n}_\r{H}$,
where $\bar{n}_\r{H} = 0.19 (1+z)^3 /\r{m}^3$ is the background value of $n_\r{H}$.
The second and third terms on the right hand side of \Eq{Eq_dxedt} represent direct and indirect ionization of hydrogen atoms from the ground state and the $n=2$ excited state respectively.
$E_\r{i} = 13.6 \r{eV}$ is the ionization energy of a ground-state hydrogen atom,
$E_\alpha = 10.2 \r{eV}$ is the energy required to excite a ground-state hydrogen atom to first excited state,
$k_\r{B}$ is the Boltzmann constant.

The Peebles factor $C$ in \Eq{Eq_dxedt} describes the probability for an excited $n=2$ hydrogen atom to transition back to the ground state before being ionized~\cite{Slatyer:2016qyl,Liu:2016cnk,Seager:1999bc},
and we compute it following \cite{dodelson2020modern},
\be
C = \frac
{\Lambda_\alpha + \Lambda_{2 \gamma}}
{\Lambda_\alpha + \Lambda_{2 \gamma} + \beta^{(2)}},
\ee
where $\Lambda_{2 \gamma} = 8.227\r{s}^{-1}$,
$\Lambda_\alpha = {H (3 E_\r{i})^3}/[{n_\r{H} (8 \pi)^2}]$ and $\beta^{(2)} = \beta \r{e}^{3 E_\r{i}/4T_\r{k}}$,
$\beta$ is the fiducial ionisation rate (see \cite{dodelson2020modern}).
{\tt 21cmFAST} assumes that hydrogen and singly ionized helium are ionized to the same degree $x_\r{e}$ until full helium reionization~\cite{Mesinger:2010ne},
which occurs at $z\sim 4$~\cite{Sokasian:2001xh} and is thus irrelevant for epochs of interest here.
The term $n_\r{H}(1+x_\r{e} + f_\r{He} + x_\r{e} f_\r{He})$ in \Eq{Eq_dTdt} represents the number density of all baryonic particles,
accounting for (from left to right)
neutral and ionized hydrogen atoms,
free electrons from ionized hydrogen,
neutral and singly ionized helium atoms,
and free electrons from singly ionized helium,
respectively.

The deposition efficiencies $f_\r{c}$ in Eqs.~(\ref{Eq_dxedt}) and (\ref{Eq_dTdt}) describe the fraction of injected energy that goes into different deposition channels,
\be
f_\r{c}
\equiv
\left[
\frac{{\rm{d}}E}
{{\rm{d}}V{\rm{d}}t}
\right]_{\rm{dep,c}}
/\left[
\frac{{\rm{d}}E}
{{\rm{d}}V{\rm{d}}t}
\right]
\label{Eq_fc_definition}
\ee
here the subscript $\r{c}$ denotes deposition channel: hydrogen ionization (ion), excitation (LyA) and IGM heating (heat),
$\left[\rd E/\rd V \rd t\right]_\r{dep, c}$ is the energy deposition rate per unit volume into channel $\r{c}$.
A systematic derivation of $f_\r{c}$ can be performed by tracking the electromagnetic cascades of DM annihilation products in the IGM,
which depends on species and primary energy of DM annihilation product, 
as well as the IGM ionization and thermal states, 
redshifts of injection and deposition, etc.
For an uniform background,
such analysis was performed in Refs.~\cite{Slatyer:2009yq,Slatyer:2012yq,Slatyer:2015kla,Slatyer:2015jla,Liu:2019bbm}.
Most recently Ref.~\cite{Sun:2023acy} studied energy deposition in an inhomogeneous background for DM decay.

However,
tracking the particle cascade in an inhomogeneous universe for annihilation process is beyond the scope of this paper.
While the energy injection rate for decaying DM is directly proportional to the simulation cell density~\cite{Sun:2023acy},
for annihilation process the inhomogeneous energy injections are contributed almost entirely by DM inside halos,
therefore for annihilating DM such analysis will most likely involve scales down to the halo size,
which is well below the simulation cell size.
The problem becomes even more complicated when taking into account the feedback from the IGM,
i.e. DM changes IGM environment, 
which in turn changes $f_\r{c}$ and thereby DM deposition itself.

For simplicity here we use the transfer function $\mathcal{T}^\r{s}_\r{c}(z, E, z')$ interpolation table provided in~\cite{Slatyer:2015kla} to track the particle energy deposition process and compute deposition efficiencies.
For a particle $\r{s}$ injected at redshift $z'$ with energy $E$,
$\mathcal{T}^\r{s}_\r{c}(z, E, z')$ gives the fraction of $E$ deposited into channel $\r{c}$ within unit $\ln a$ time step around redshift $z$.
It can be shown analytically~\cite{Slatyer:2012yq,Cang:2021owu,Lopez-Honorez:2013cua} that for a generic particle injection process,
the relevant energy deposition $\left[\rd E/\rd V \rd t\right]_\r{dep, c}$ is given by,
\be
\begin{aligned}
\left[
\frac{{\rm{d}}E}
{{\rm{d}}V{\rm{d}}t}
\right]_{\rm{dep,c}}
(z)
&=
(1+z)^3 H(z)
\int_z^{\infty}
\frac
{\r{d} z'}
{(1+z')^4 H(z')}
\\
&\times
\sum_{\r{s} = \gamma, e^{\pm}}
\int \r{d}E
\ 
E
\mathcal{T}^\r{s}_{\r{c}}(z,E,z')
\mathcal{R}^\r{s}(E,z')
,
\end{aligned}
\label{Dep_EFF_EQ}
\ee
where the summation is over photon, electron and positron,
which are the only stable Standard Model particles with sufficient electromagnetic interaction with the IGM.
The factor $\mathcal{R}^\r{s}$ describes the number of $\r{s}$ particles injected per unit energy, time and volume.
For simplicity here we will consider DM annihilation into an electron-positron pair ($\chi \chi \to \r{e}^-\r{e}^+$),
for which $\mathcal{R}^\r{s}$ takes the form,
\be
\mathcal{R}^{\r{e}^-}
=
\mathcal{R}^{\r{e}^+}
=
\frac{B \sv}{2 m_\chi^2}
\Omega_\r{c}^2 \rho_\r{cr}^2
(1+z)^6
\delta(E-m_\chi).
\label{Eq_IFunction_full}
\ee

Inserting this equation into \Eq{Dep_EFF_EQ},
from \Eq{Eq_fc_definition} one can then find the expression for the deposition efficiency,
\be
\begin{aligned}
f_\r{c}
(z, m_\chi)
& = 
\frac{H(z)}{2 \bar{B}(z) (1+z)^3}
\sum_{\r{s} = e^{\pm}}
\int_z^{\infty} \frac{\rd z'}{H(z')}
(1+z')^2
\\
&\times \bar{B}(z') \mathcal{T}^\r{s}_\r{c}
(z,m_\chi,z'),
\end{aligned}
\label{Eq_fc_analytic}
\ee
which now depends only on the DM mass $m_\chi$ and the deposition redshift and can be pre-computed and used as an interpolation table for computational convenience.
Due to challenges outlined out in previous texts,
$\mathcal{T}^\r{s}_\r{c}(z, E, z')$ does not resolve energy deposition down to the scale of the simulation cell or halo size and thus does not reflect density fluctuation,
thus in \Eq{Eq_fc_analytic} we use the spatially averaged boost factor $\bar{B}$ rather than its spatial-dependent counterpart.

We divide our calculation of Eqs.~(\ref{Eq_dxedt}) and (\ref{Eq_dTdt}) into two stages:
at high redshifts ($z > 60$),
we solve the recombination history using our modified {\tt HyRec} package~\cite{Ali-Haimoud:2010hou,Lee:2020obi}.
As this is well before the Stelliferous Era,
we ignore stellar radiation in Eqs.~(\ref{Eq_dxedt}, \ref{Eq_dTdt}) during this epoch. In the lower redshift stage ($z \leq 60$),
we perform the full {\tt 21cmFAST} simulation using the initial conditions for $x_\r{e}$ and $T_\r{k}$ set by high-redshift evolution.

\section{Simulation Results}
\label{juhhdyf83_dhfdsgvt}

We showcase the effect of inhomogeneous DM annihilation with the following three simulation settings,
\begin{itemize}

\item {\tt Inhomogeneous Boost (IHM):} 
Our main simulation assuming the inhomogeneous injection boost factor and energy deposition formalism detailed in \Sec{Section_2} and the background astrophysics described in \Sec{juhhdyf83_dhfdgvt}.
We consider the annihilation channel $\chi \chi \to \r{e}^+ \r{e}^-$ with annihilation rate and mass set to $\sv/\mdm = 10^{-27} \r{cm^3 s^{-1} GeV^{-1}}$ and 100 MeV respectively,
this injection rate roughly corresponds to the current CMB constraints from {\it Planck}~\cite{Planck:2018vyg}.

\item {\tt Homogeneous Boost (HMG):}
Same as the {\tt IHM} case except that instead of the inhomogeneous boost factor,
we use its global average value $\bar{B}$.

\item {\tt Fiducial:} 
Simulation for the fiducial astrophysical background detailed in \Sec{juhhdyf83_dhfdgvt} in absence of DM injection ($\sv/\mdm = 0$).
\end{itemize}

\begin{figure*}[htp]
\centering
\includegraphics[width=20.5cm]{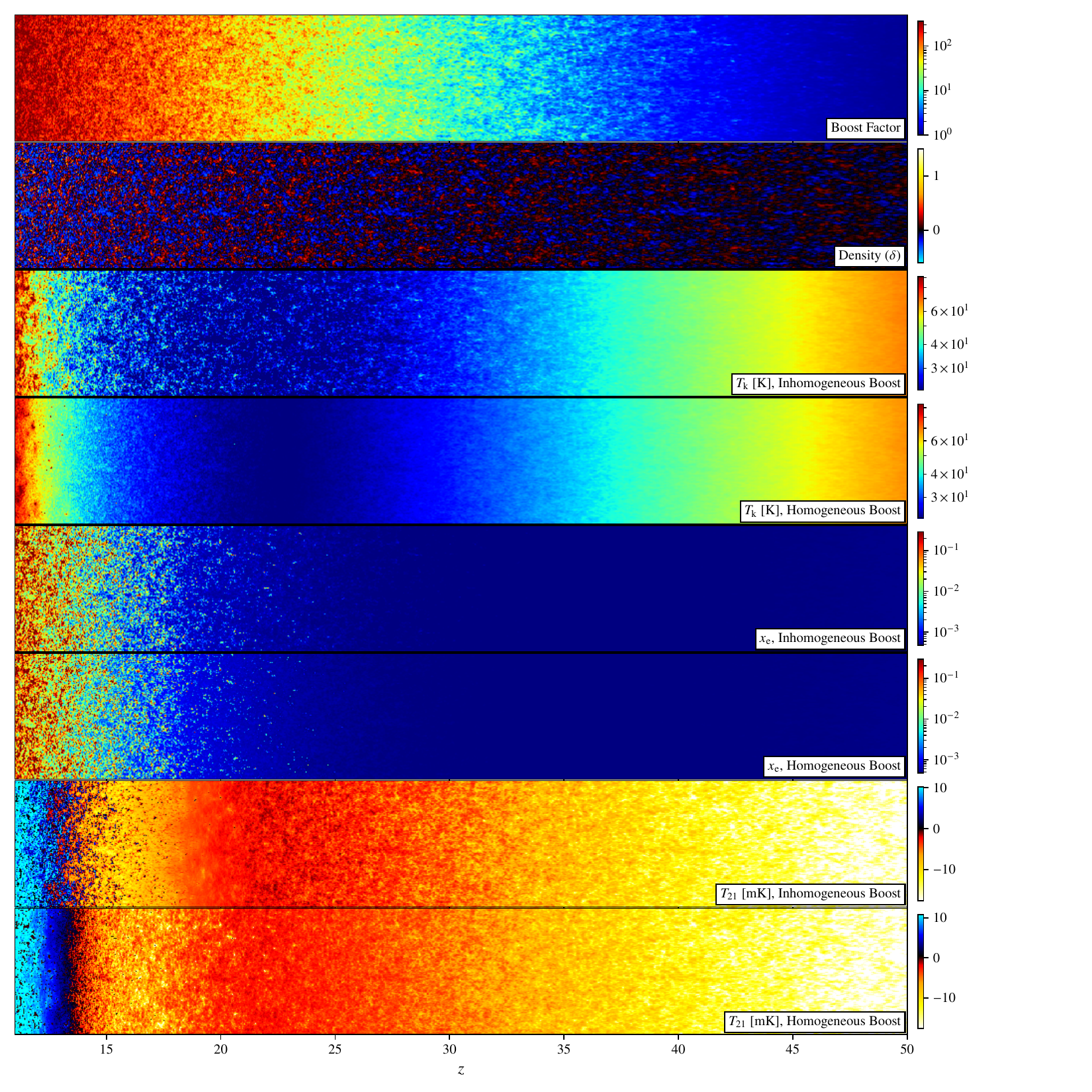}
\caption{
Lightcone simulations of the inhomogeneous boost factor (top) and the density field ($\delta$).%($\rho_\r{m}/\bar{\rho}_\r{m}$).
The third to eighth panels visualize the evolution of gas temperature $\tk$, ionisation fraction $\xe$ and 21-cm temperature $\tb$ in presence of inhomogeneous/homogeneous boost factor,
both using $\left<\sigma v\right>/m_\chi = 10^{-27} {\rm cm^3 s^{-1} GeV^{-1}}$ and $m_\chi = 100\ \r{MeV}$.
The panels labeled with {\it Inhomogeneous Boost} correspond to the scenario in which DM annihilation products have relatively low energy and are therefore absorbed locally. 
If DM injects high-energy particles that have a long mean free path before absorption,
the effect of inhomogeneous boost factor would be washed out.
This corresponds to the panels labeled with {\it homogeneous boost},
for which the boost factor is added uniformly using its spatially averaged value.
Note that the growth and fluctuation of the boost factor trace those of the density field,
and the panels with inhomogeneous boost factor exhibit more fluctuations than those with homogeneous boost factor.
The power spectrum shown in \Fig{Fig_Power_T21} provides more quantitative comparison of these inhomogeneities.
}
\label{Fig_LightCone_Plot}
\end{figure*}

\begin{figure*}[htp]
\centering
\subfigbottomskip=-500pt
\includegraphics[width=18cm]{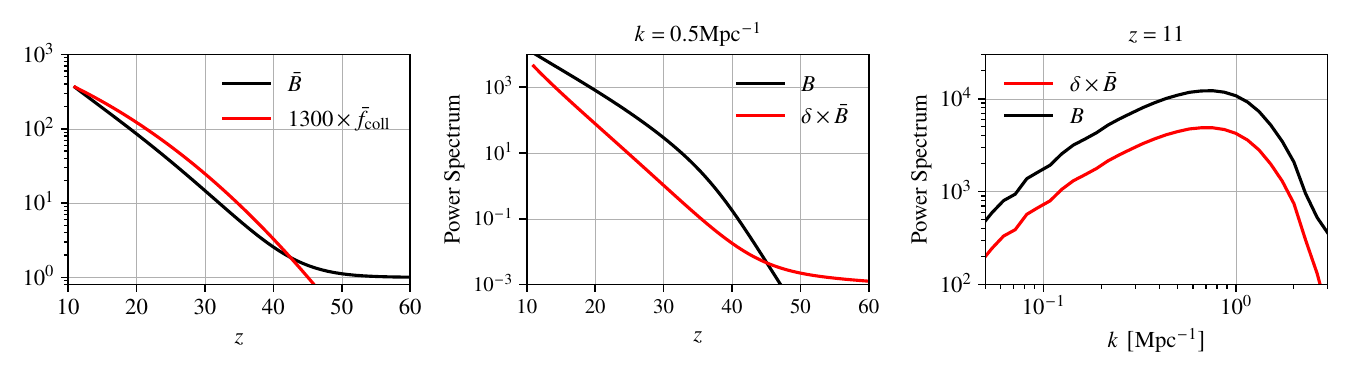}
\caption{
{\it Left}--global average of boost factor $B$ (black) and collapse fraction $f_\r{coll}$ (red).
{\it Middle and right}--power spectrum for $B$ (black) and density contrast (red, scaled by the mean boost factor $\bar{B}$).
}
\label{e2ftssaasadwu}
\end{figure*}

\begin{figure}[b]
\centering
\includegraphics[width=9cm]{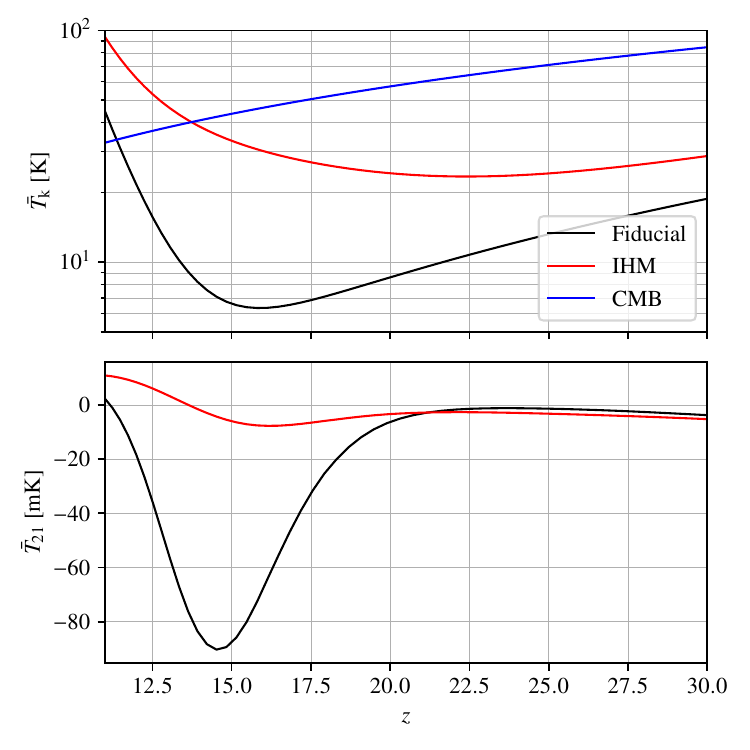}
\caption
{
Evolution of spatially averaged $T_\r{k}$ (top) and $T_{21}$ (bottom).
The black and red lines represents the {\tt Fiducial} and {\tt IHM} simulations respectively,
results for {\tt HMG} are found to be the same as that in {\tt IHM} simulation and are thus not shown here.
The blue line on top panel shows the CMB temperature.
The legend applies to both panels.
}
\label{Fig_Global_QUantities}
\end{figure}

\begin{figure*}[htp]
\centering
\includegraphics[width=18cm]{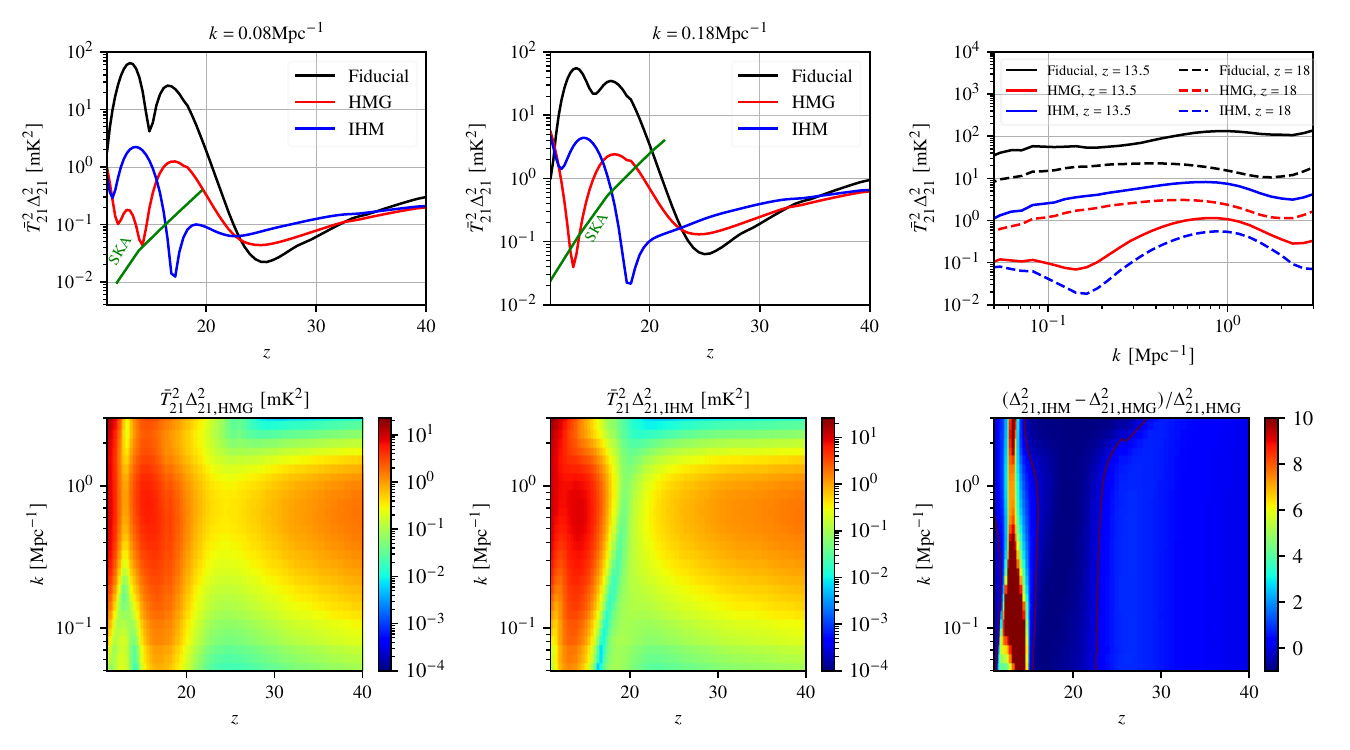}
\caption
{
{\it Top}:
the 21-cm power spectrum $\bar{T}^2_{21}\Delta_{21}^2$ at several characteristic scales and redshifts.
The black, red and blue lines represent the {\tt Fiducial}, {\tt HMG} and {\tt IHM} simulations respectively.
The left and middle panels show $\bar{T}^2_{21}\Delta_{21}^2$ at $k = 0.08 \r{Mpc}^{-1}$ and $k = 0.18 \r{Mpc}^{-1}$ respectively,
and the right panel shows results for $z=13.5$ (solid) and $z=18$ (dashed).
Both {\tt IHM} and {\tt HMG} simulations assume the same DM injection parameters of $\left<\sigma v\right>/m_\chi = 10^{-27} {\rm cm^3 s^{-1} GeV^{-1}}$ and $m_\chi = 100\ \r{MeV}$,
in {\tt HMG} simulation $\bar{T}^2_{21}\Delta_{21}^2$ can be enhanced by orders of magnitude compared to that of {\tt IHM}.
Such difference can potentially be detected by the SKA telescope with 2000 hours of observation time (green solid curve)~\cite{Sitwell:2013fpa}.
{\it Bottom}:
$\bar{T}^2_{21}\Delta_{21}^2$ at different scales and redshifts,
the left and middle panels correspond to the {\tt HMG} and {\tt IHM} simulations,
respectively.
The right panel shows their relative difference,
in regions outside the black contour line,
$\bar{T}^2_{21}\Delta^2_{21}$ for the {\tt IHM} simulation is enhanced relative to the {\tt HMG} case.
For visual illustration,
the color bar is truncated around 10, 
and the actual relative difference can be much higher in some regions.
}
\label{Fig_Power_T21}
\end{figure*}

All our simulations are performed with $300^3$ resolution and a box length of 500 comoving Mpc.
Unless otherwise specified,
all length scales in the following text are expressed in comoving units.
The energy loss process of DM annihilation products is dependent on IGM ionization and thermal states,
and the energy deposition transfer function we used from~\cite{Slatyer:2015kla} is truncated at $z<11$ due to uncertainties in the IGM ionization and temperature at these lower redshifts,
thus we also truncate our simulation at $z=11$.
If DM annihilation products have a long mean free path before absorption,
the heating and ionization rates will remain widely dispersed despite the inhomogeneity in injection rate. 
The {\tt HMG} simulation can be seen as a representation of this scenario, 
and to a large extent aligns with the situation in previous analyses in~\cite{Taylor:2002zd,Liu:2016cnk,Lopez-Honorez:2016sur,Short:2019twc,Huetsi:2009ex,Poulin:2015pna,Diamanti:2013bia,Natarajan:2010dc,Valdes:2012zv}.
In contrast,
as represented by the {\tt IHM} simulation, 
if the annihilation products have a short absorption length, 
which is typically the case if they are injected below electroweak energy scale or if they consist mainly of electrons~\cite{Slatyer:2012yq,Slatyer:2009yq,Slatyer:2015kla}, 
the injected energy would be deposited locally, 
and the corresponding ionization and heating rates would be inhomogeneous.

In the {\tt IHM} simulation,
since $f_\r{c}$ is homogeneous in our prescription,
the inhomogeneity of deposition rate depends only on the inhomogeneity of the local injection rate,
which is a good approximation if energy deposition is instantaneous such that $\mathcal{T}^\r{s}_\r{c}(z, E, z')$ approaches $\delta (z-z')$.
However for non-local deposition,
the deposited energy should also receive contributions from particles propagated from neighboring simulation cells and this can smear the inhomogeneity in deposition rate and induce inhomogeneity in $f_\r{c}$ itself.
Our prescription does not capture these effects and we reserve relevant studies for our future work.

We focus on a DM mass of 100 MeV,
which maximizes the deposition efficiencies into both heating and ionization~\cite{Slatyer:2015jla, Slatyer:2015kla}. 
For other masses,
the effects of DM annihilation are expected to be less pronounced.
Above the 3 keV kinetic energy scale, 
electrons and positrons predominantly lose energy through inverse Compton scattering (ICS)~\cite{Slatyer:2015jla, Slatyer:2015kla}. 
For our DM mass of 100 MeV,
ICS of the injected electrons and positrons primarily produces photons in the $100$–$10^4$~eV range, 
which efficiently ionize hydrogen over length scales much shorter than the Hubble scale~\cite{Slatyer:2015jla, Evoli:2012zz}, 
leading to spatially inhomogeneous ionization and heating. 
At higher energies ($\gtrsim$~GeV), 
ICS-produced photons are much more energetic and interact less efficiently with the IGM,
resulting in longer absorption lengths~\cite{Evoli:2012zz} and reduced inhomogeneity in ionization and heating.

\Fig{Fig_LightCone_Plot} shows the inhomogeneous lightcone evolutions for the boost factor,
density contrast,
$x_{\rm{e}},\ T_{\rm{k}}$ and $T_{21}$,
respectively,
along with comparisons between the {\tt IHM} and {\tt HMG} simulations.
From the upper two panels,
it can be seen that the boost factor $B$ exhibits distinctive inhomogeneity patterns that closely trace those in density contrast $\delta$.
This is further demonstrated quantitatively in the middle and right panels of \Fig{e2ftssaasadwu},
where we show that after normalization,
$B$ and $\delta$ share remarkably similar power spectrum.
The amplitude of $B$ also traces the density fluctuation level,
as can be inferred from the top two panels of \Fig{Fig_LightCone_Plot}.
At high redshifts ($z > 50$) when the inhomogeneity in matter distribution is negligible,
$B$ takes unity and increases with the growth of density fluctuation.
Since the collapse fraction $f_\r{coll}$ indirectly reflects the density fluctuation level,
this can also be seen in the left panel of \Fig{e2ftssaasadwu},
which shows that $B$ follows a growth history similar to that of $f_\r{coll}$.
By redshift $z = 11$ when about $27\%$ of matter has collapsed into halos,
DM annihilation rate is boosted by roughly a factor of 350 compared to the uniform background.

\begin{figure*}[htp]
\centering
\subfigbottomskip=-500pt
\includegraphics[width=18cm]{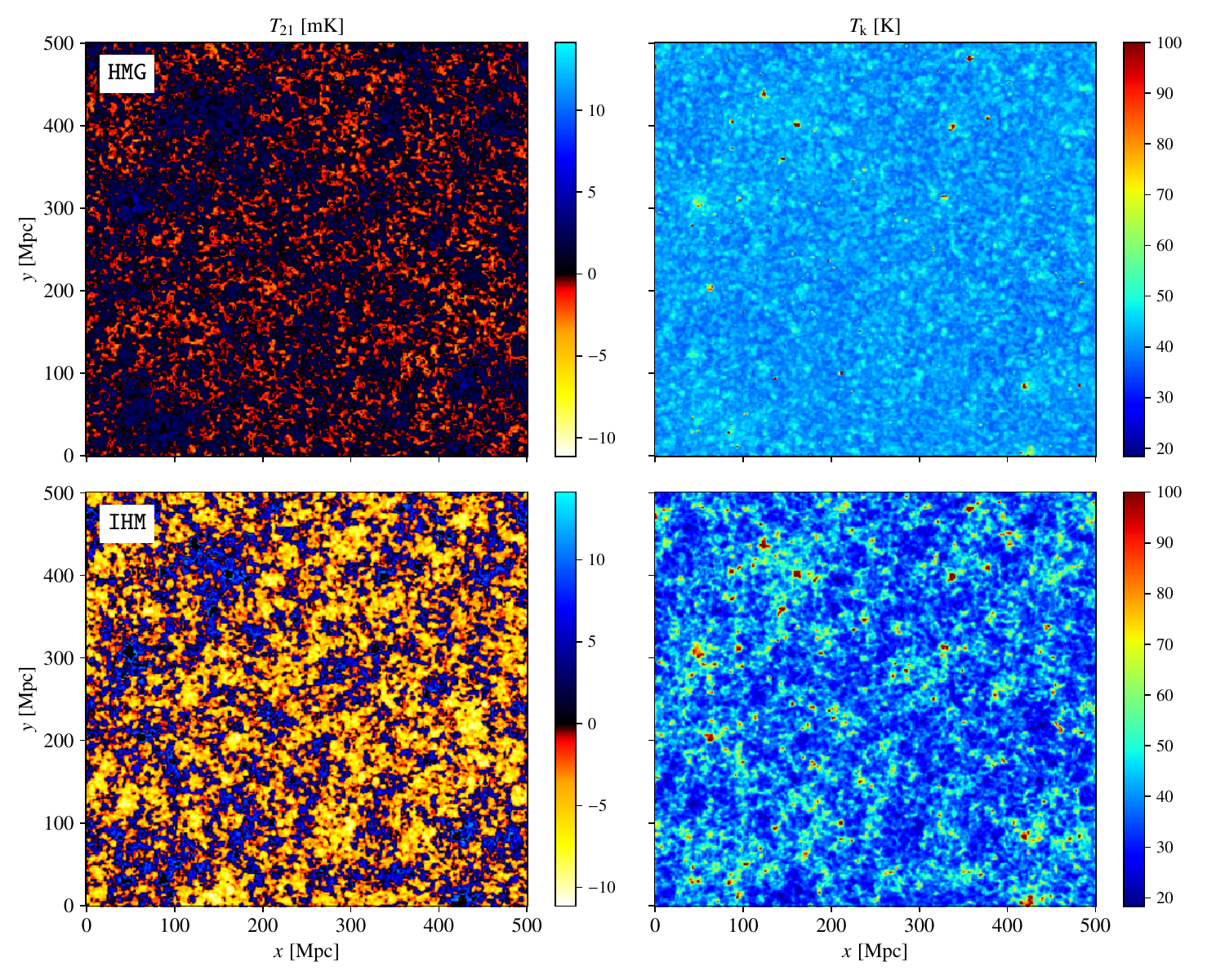}
\caption{
Spatial slice of our simulation lightcones for $T_{21}$ (left) and $T_\r{k}$ (right) at $z = 13.5$,
corresponding to the redshift when the enhancement of 21-cm power spectrum in {\tt IHM} (bottom) relative to the {\tt HMG} (top) simulation is maximum.
To highlight the fluctuation patterns,
in $T_\r{k}$ panels we truncate the color bar for $T_\r{k} > 100$ K.
}
\label{Fig_Coeval_Slice}
\end{figure*}

\Fig{Fig_Global_QUantities} shows the evolutions of global $T_\r{k}$ and $T_{21}$ from {\tt Fiducial} and {\tt IHM} simulations.
All global quantities ($\bar{T}_\r{k}$, $\bar{T}_{21}$ and $\bar{x}_\r{H}$) in the {\tt HMG} simulation are found to be nearly identical to that in {\tt IHM} scenario and are thus not shown independently.
For our chosen annihilation rate,
ionizing radiations from DM is subdominant to that from astrophysical sources,
and we find that the neutral fraction $x_\r{H}$ from both the {\tt IHM} and {\tt HMG} simulations remain practically the same as that in the {\tt Fiducial} case,
thus in our annihilation scenario the impact of DM on 21-cm signal is driven mainly by the heating term.
As can be seen from the top panel,
the heating from DM starts much earlier ($z > 30$) than the astrophysical sources and lead to suppressed absorption ($T_{21} < 0$) signal.
By redshift 13.8,
DM heats the IGM temperature to above that of CMB,
leading to the transition of 21-cm signal from absorption to emission ($T_{21} > 0$),
whereas in the {\tt Fiducial} simulation the transition occurs only later at $z=11$.
%\ck{added paragraph, remove this text after editing referee reply}

In \Fig{Fig_Power_T21} we present comparisons of 21-cm power spectrum from our simulations.
We highlight the power spectrum at $k = 0.08 \r{Mpc^{-1}}$ and $k = 0.18 \r{Mpc^{-1}}$,
corresponding to scales that are large enough for efficient foreground removal and yet small enough for experiments to achieve high signal to noise ratio~\cite{Mesinger:2013nua,Lidz:2007az,Dillon:2013rfa,Pober:2013ig}.
We also include the forecasted $1\sigma$ power spectrum sensitivity for the SKA (Square Kilometer Array) telescope computed in Refs.~\cite{Sitwell:2013fpa,Mesinger:2013nua},
which assumed 2000 hours of observation time and an observational strategy carefully chosen to minimize thermal noise.

As can be seen from \Fig{Fig_LightCone_Plot},
compared to the {\tt HMG} simulation,
$T_\r{k}$ and $T_\r{21}$ both display remarkably different inhomogeneity in the {\tt IHM} simulation.
For the redshift window $z \in [11,\ 16]$ in particular,
inhomogeneity levels are significantly enhanced in {\tt IHM} simulation due to fluctuations in the DM heating rate.
In comparison,
$x_\r{e}$ exhibits only moderately enhanced fluctuations within the redshift range $z \sim 18$–28. 
At lower redshifts,
ionization from astrophysical sources dominates,
leading to similar $x_\r{e}$ evolution in both the {\tt IHM} and {\tt HMG} simulations.
The 21-cm power spectrum in \Fig{Fig_Power_T21} provides quantitative comparisons of these fluctuations.
For the {\tt HMG} simulation, 
the change in $\bar{T}^2_{21}\Delta^2_{21}$ relative to the {\tt Fiducial} setting is largely driven by the difference in 21-cm amplitude:
qualitatively similar to that in \cite{Lopez-Honorez:2016sur},
in this case heating from DM weakens 21-cm absorption signal and thereby suppressing the 21-cm power spectrum amplitude.
For the {\tt IHM} scenario,
$\bar{T}^2_{21}\Delta^2_{21}$ is also affected by spatial variation of $T_{21}$ induced by inhomogeneous DM heating/ionization.
As shown in \Fig{Fig_Power_T21},
accounting for inhomogeneities in annihilation can enhance $\bar{T}^2_{21}\Delta^{2}_{21}$ by up to a factor of 130 for $11 \lesssim z\lesssim 16$,
and such enhancement is potentially detectable with the SKA telescope~\cite{Sitwell:2013fpa}.
The enhancement in 21-cm power spectrum is maximized at around $z=13.5$,
\Fig{Fig_Coeval_Slice} shows the spatial slices of $T_{21}$ (left) and $T_\r{k}$ (right) fields at this redshift,
and it can be seen that while {\tt HMG} (top) and {\tt IHM} (bottom) simulations share the same global values for $T_{21}$ and $T_\r{k}$,
both observables shows strikingly different fluctuation features.

While \Fig{Fig_LightCone_Plot} shows that both gas temperature and ionization fraction display more fluctuation features in the inhomogeneous annihilation than in the homogeneous case,
in \Fig{Fig_Power_T21} (e.g., upper and lower right panels) one sees that inhomogeneous annihilation in fact suppresses 21-cm power spectrum at $16 \lesssim z \lesssim 23$ when the 21-cm signal is in absorption ($T_\r{s} < T_\gamma$).
For our chosen DM mass and annihilation channel,
energy deposition at a given cell is contributed mostly by nearby sources,
and from discussions layed out in \Sec{Section_2},
it can be inferred that in inhomogeneous annihilation,
overdense cells ($\delta_\r{b} > 0$) receive more DM heating than underdense regions ($\delta_\r{b} < 0$) and thus have higher kinetic and spin temperatures.
It can be inferred from \Eq{Eq_T21_definition} that this effect alone suppresses the 21-cm absorption signal amplitude,
whereas the density fluctuation term $1+\delta_\r{b}$ alone amplifies 21-cm signal in overdense regions.
For redshifts under discussion,
fluctuations in the 21-cm signal are dominated by inhomogeneous density field,
and the anti-correlation between effects of DM heating and density field suggests that heating from inhomogeneous DM annihilation suppresses 21-cm fluctuations and thereby lowers the 21-cm power spectrum.
A similar argument can also be applied to show that when the 21-cm signal is in absorption,
ionization from inhomogeneous annihilation also suppresses the 21-cm power spectrum.

At even higher redshifts ($z\gtrsim 23$),
\Fig{Fig_Power_T21} shows that 21-cm power spectrum is again enhanced in the {\tt IHM} simulation compared to {\tt HMG} case.
At these redshifts,
heating from both astrophysical sources and DM annihilation are less pronounced and the gas kinetic temperature is below CMB temperature $T_\gamma$.
In overdense regions,
due to higher energy deposition and consequently stronger Wouthuysen-Field coupling (see \Eq{Eq_J_LyA_DM}),
spin temperature would be cooled closer to kinetic temperature,
this effect alone enhances 21-cm absorption signal and is positively correlated with the impact of density field,
thus at these highest redshifts during which 21-cm fluctuation is mainly driven by density field,
21-cm power spectrum is enhanced by inhomogeneous annihilation.

\section{Discussions}
\label{juhhdsayf88hhfe3_dhfdsgvt}

Particles injected from DM annihilation events can heat up and ionize the intergalactic medium and change the 21-cm signal from neutral hydrogen during the cosmic dawn.
At low redshifts,
the growth of structures can significantly boost DM annihilation rate relative to the uniform background.
This paper examines the inhomogeneity in DM annihilation boost factor and its impact on 21-cm brightness temperature power spectrum $\bar{T}^2_{21}\Delta^2_{21}$.
Building on the {\tt 21cmFAST} simulation framework and the conditional halo mass function,
we obtain the lightcone evolution and the power spectrum for the inhomogeneous boost factor.
We showcase the effect for $\chi \chi \to \r{e}^- \r{e}^+$ annihilation channel with the DM mass and annihilation rate of 100 MeV and $\sv/\mdm = 10^{-27} \r{cm^3 s^{-1} GeV^{-1}}$ respectively,
and our results show that compared to the case with homogeneous boost factor,
which approximate long propagation length for DM annihilation products that tends to wash out footprints of inhomogeneous injection,
the inhomogeneous boost factor can induce distinctively different fluctuation features in 21-cm signal in the redshift window of $11 \lesssim z \lesssim 16$,
and the corresponding 21-cm power spectrum $\bar{T}^2_{21}\Delta^2_{21}$ can be enhanced by up to a factor of 130.
Such features could potentially be detected by the SKA telescope.

We focused on a DM mass of 100 MeV in this work,
however we expect our conclusion to hold across a wide DM mass range of [0.51 MeV, 10 TeV] for $\r{e}^\pm$ annihilation channel.
For inhomogeneous annihilation to impact 21-cm fluctuation compared with the homogeneous model,
DM annihilation products must deposit the majority of their energy over distances shorter than the scale of interest.
Within the redshifts explored here,
it was shown in \cite{Sun:2023acy} that for $\r{e^\pm}$ with kinetic energies in either the $10^2$–$10^5$~eV or $10^7$–$10^{12}$~eV ranges,
the majority of the injected energy is deposited before they can travel the distance of up to 4.6 Mpc,
which is comparable to our simulation resolution of 1.7 Mpc and well below the 300 Mpc scale accessible to SKA \cite{Koopmans:2015sua}.
Even within the intermediate energy range ($10^5-10^7$~eV),
\cite{Sun:2023acy} argued that realistic strength of intergalactic magnetic fields can potentially confine $\r{e^\pm}$ propagation to proper distances of $\lesssim 0.05$ Mpc,
which is much smaller than our simulation resolution for redshifts of interest here.
We expect that effects of inhomogeneous annihilation will be less significant for annihilation into other standard model particles (e.g., photons and neutrinos),
which in general interact less efficiently with IGM \cite{Sun:2023acy,Evoli:2012zz} and have much longer absorption length.

We compute the energy deposition fractions into different absorption channels by tracking the particle cascade using the transfer function provided in~\cite{Slatyer:2015kla}.
Our prescription does not capture the inhomogeneous energy propagation from neighboring simulation cells which can smear the inhomogeneity in local deposition rate.
A more comprehensive analysis of the energy deposition process would necessitate a detailed study of particle cascade and propagation in an inhomogeneous background as well as the feedback from IGM thermal and ionization states.
Furthermore,
we showcased the detectability of DM signature with SKA by comparing SKA sensitivity with the power spectrum for our illustrative model,
in which we fixed astrophysical and DM parameters.
A more robust and quantitative assessment of detectability would require a more detailed SKA forecast analysis with varied astrophysical and DM parameters.
However these analyses are beyond the scope of this paper and we reserve such investigations for future works.

\medskip
{\bf Acknowledgements}\\ 
We thank Andrei Mesinger, Yuxiang Qin and Steven Murray for their helpful communications. This work is supported by the National Natural Science Foundation of China (grant No.~12275278), the National Research Foundation with grant No.~150580, and the research program ``New Insights into Astrophysics and Cosmology with Theoretical Models Confronting Observational Data'' of the National Institute for Theoretical and Computational Sciences of South Africa. 

\section*{Data Availability}
The data that support the findings of this article are not publicly available upon publication because it is not technically feasible and/or the cost of preparing, depositing, and hosting the data would be prohibitive within the terms of this research project. 
The data are available from the authors upon reasonable request.
\twocolumngrid
\bibliography{main}

@article{Planck:2018vyg,
    author = "Aghanim, N. and others",
    collaboration = "Planck",
    title = "{Planck 2018 results. VI. Cosmological parameters}",
    eprint = "1807.06209",
    archivePrefix = "arXiv",
    primaryClass = "astro-ph.CO",
    doi = "10.1051/0004-6361/201833910",
    journal = "Astron. Astrophys.",
    volume = "641",
    pages = "A6",
    year = "2020",
    note = "[Erratum: Astron.Astrophys. 652, C4 (2021)]"
}

@article{Ziparo:2022fnc,
    author = "Ziparo, Francesco and Gallerani, Simona and Ferrara, Andrea and Vito, Fabio",
    title = "{Cosmic radiation backgrounds from primordial black holes}",
    eprint = "2209.09907",
    archivePrefix = "arXiv",
    primaryClass = "astro-ph.CO",
    doi = "10.1093/mnras/stac2705",
    journal = "Mon. Not. Roy. Astron. Soc.",
    volume = "517",
    number = "1",
    pages = "1086--1097",
    year = "2022"
}

@article{Navarro:1996gj,
    author = "Navarro, Julio F. and Frenk, Carlos S. and White, Simon D. M.",
    title = "{A Universal density profile from hierarchical clustering}",
    eprint = "astro-ph/9611107",
    archivePrefix = "arXiv",
    doi = "10.1086/304888",
    journal = "Astrophys. J.",
    volume = "490",
    pages = "493--508",
    year = "1997"
}

@article{Lacey:1993iv,
    author = "Lacey, Cedric G. and Cole, Shaun",
    title = "{Merger rates in hierarchical models of galaxy formation}",
    journal = "Mon. Not. Roy. Astron. Soc.",
    volume = "262",
    pages = "627--649",
    year = "1993"
}

@article{Somerville:1997df,
    author = "Somerville, Rachel S. and Kolatt, Tsafrir S.",
    title = "{How to plant a merger tree}",
    eprint = "astro-ph/9711080",
    archivePrefix = "arXiv",
    doi = "10.1046/j.1365-8711.1999.02154.x",
    journal = "Mon. Not. Roy. Astron. Soc.",
    volume = "305",
    pages = "1--14",
    year = "1999"
}

@article{Park:2018ljd,
    author = "Park, Jaehong and Mesinger, Andrei and Greig, Bradley and Gillet, Nicolas",
    title = "{Inferring the astrophysics of reionization and cosmic dawn from galaxy luminosity functions and the 21-cm signal}",
    eprint = "1809.08995",
    archivePrefix = "arXiv",
    primaryClass = "astro-ph.GA",
    doi = "10.1093/mnras/stz032",
    journal = "Mon. Not. Roy. Astron. Soc.",
    volume = "484",
    number = "1",
    pages = "933--949",
    year = "2019"
}

@article{Mesinger:2010ne,
    author = "Mesinger, Andrei and Furlanetto, Steven and Cen, Renyue",
    title = "{21cmFAST: A Fast, Semi-Numerical Simulation of the High-Redshift 21-cm Signal}",
    eprint = "1003.3878",
    archivePrefix = "arXiv",
    primaryClass = "astro-ph.CO",
    doi = "10.1111/j.1365-2966.2010.17731.x",
    journal = "Mon. Not. Roy. Astron. Soc.",
    volume = "411",
    pages = "955",
    year = "2011"
}

@article{Pritchard:2011xb,
    author = "Pritchard, Jonathan R. and Loeb, Abraham",
    title = "{21-cm cosmology}",
    eprint = "1109.6012",
    archivePrefix = "arXiv",
    primaryClass = "astro-ph.CO",
    doi = "10.1088/0034-4885/75/8/086901",
    journal = "Rept. Prog. Phys.",
    volume = "75",
    pages = "086901",
    year = "2012"
}

@article{Furlanetto:2006jb,
    author = "Furlanetto, Steven and Oh, S. Peng and Briggs, Frank",
    title = "{Cosmology at Low Frequencies: The 21 cm Transition and the High-Redshift Universe}",
    eprint = "astro-ph/0608032",
    archivePrefix = "arXiv",
    doi = "10.1016/j.physrep.2006.08.002",
    journal = "Phys. Rept.",
    volume = "433",
    pages = "181--301",
    year = "2006"
}

@article{Cooray:2002dia,
    author = "Cooray, Asantha and Sheth, Ravi K.",
    title = "{Halo Models of Large Scale Structure}",
    eprint = "astro-ph/0206508",
    archivePrefix = "arXiv",
    reportNumber = "FERMILAB-PUB-02-284-A",
    doi = "10.1016/S0370-1573(02)00276-4",
    journal = "Phys. Rept.",
    volume = "372",
    pages = "1--129",
    year = "2002"
}

@article{Bouwens:2014fua,
    author = "Bouwens, R. J. and others",
    title = "{UV Luminosity Functions at redshifts $z \sim$4 to $z \sim$10: 10000 Galaxies from HST Legacy Fields}",
    eprint = "1403.4295",
    archivePrefix = "arXiv",
    primaryClass = "astro-ph.CO",
    doi = "10.1088/0004-637X/803/1/34",
    journal = "Astrophys. J.",
    volume = "803",
    number = "1",
    pages = "34",
    year = "2015"
}

@article{Bouwens:2015vha,
    author = "Bouwens, R. J. and Illingworth, G. D. and Oesch, P. A. and Caruana, J. and Holwerda, B. and Smit, R. and Wilkins, S.",
    title = "{Reionization after Planck: The Derived Growth of the Cosmic Ionizing Emissivity now matches the Growth of the Galaxy UV Luminosity Density}",
    eprint = "1503.08228",
    archivePrefix = "arXiv",
    primaryClass = "astro-ph.CO",
    doi = "10.1088/0004-637X/811/2/140",
    journal = "Astrophys. J.",
    volume = "811",
    number = "2",
    pages = "140",
    year = "2015"
}

@article{Oesch_2018,
doi = {10.3847/1538-4357/aab03f},
url = {https://dx.doi.org/10.3847/1538-4357/aab03f},
year = {2018},
month = {mar},
publisher = {The American Astronomical Society},
volume = {855},
number = {2},
pages = {105},
author = {P. A. Oesch and R. J. Bouwens and G. D. Illingworth and I. Labbé and M. Stefanon},
title = {The Dearth of z ∼ 10 Galaxies in All HST Legacy Fields—The Rapid Evolution of the Galaxy Population in the First 500 Myr*},
journal = {The Astrophysical Journal}
}

@article{Planck:2016mks,
    author = "Adam, R. and others",
    collaboration = "Planck",
    title = "{Planck intermediate results. XLVII. Planck constraints on reionization history}",
    eprint = "1605.03507",
    archivePrefix = "arXiv",
    primaryClass = "astro-ph.CO",
    doi = "10.1051/0004-6361/201628897",
    journal = "Astron. Astrophys.",
    volume = "596",
    pages = "A108",
    year = "2016"
}

@article{McGreer:2014qwa,
    author = "McGreer, Ian and Mesinger, Andrei and D'Odorico, Valentina",
    title = "{Model-independent evidence in favour of an end to reionization by $z \approx$ 6}",
    eprint = "1411.5375",
    archivePrefix = "arXiv",
    primaryClass = "astro-ph.CO",
    doi = "10.1093/mnras/stu2449",
    journal = "Mon. Not. Roy. Astron. Soc.",
    volume = "447",
    number = "1",
    pages = "499--505",
    year = "2015"
}

@article{Barkana:2000fd,
    author = "Barkana, Rennan and Loeb, Abraham",
    title = "{In the beginning: The First sources of light and the reionization of the Universe}",
    eprint = "astro-ph/0010468",
    archivePrefix = "arXiv",
    doi = "10.1016/S0370-1573(01)00019-9",
    journal = "Phys. Rept.",
    volume = "349",
    pages = "125--238",
    year = "2001"
}

@article{Liu:2016cnk,
    author = "Liu, Hongwan and Slatyer, Tracy R. and Zavala, Jes\'us",
    title = "{Contributions to cosmic reionization from dark matter annihilation and decay}",
    eprint = "1604.02457",
    archivePrefix = "arXiv",
    primaryClass = "astro-ph.CO",
    reportNumber = "MIT-CTP-4797",
    doi = "10.1103/PhysRevD.94.063507",
    journal = "Phys. Rev. D",
    volume = "94",
    number = "6",
    pages = "063507",
    year = "2016"
}

@article{Lopez-Honorez:2016sur,
    author = "Lopez-Honorez, Laura and Mena, Olga and Molin\'e, \'Angeles and Palomares-Ruiz, Sergio and Vincent, Aaron C.",
    title = "{The 21 cm signal and the interplay between dark matter annihilations and astrophysical processes}",
    eprint = "1603.06795",
    archivePrefix = "arXiv",
    primaryClass = "astro-ph.CO",
    reportNumber = "CFTP-16-007, IFIC-16-16, IPPP-16-20",
    doi = "10.1088/1475-7516/2016/08/004",
    journal = "JCAP",
    volume = "08",
    pages = "004",
    year = "2016"
}

@article{Taylor:2002zd,
    author = "Taylor, James E. and Silk, Joseph",
    title = "{The Clumpiness of cold dark matter: Implications for the annihilation signal}",
    eprint = "astro-ph/0207299",
    archivePrefix = "arXiv",
    doi = "10.1046/j.1365-8711.2003.06201.x",
    journal = "Mon. Not. Roy. Astron. Soc.",
    volume = "339",
    pages = "505",
    year = "2003"
}

@article{Short:2019twc,
    author = "Short, Katie and Bernal, Jos\'e Luis and Raccanelli, Alvise and Verde, Licia and Chluba, Jens",
    title = "{Enlightening the dark ages with dark matter}",
    eprint = "1912.07409",
    archivePrefix = "arXiv",
    primaryClass = "astro-ph.CO",
    doi = "10.1088/1475-7516/2020/07/020",
    journal = "JCAP",
    volume = "07",
    pages = "020",
    year = "2020"
}

@article{Huetsi:2009ex,
    author = "Huetsi, Gert and Hektor, Andi and Raidal, Martti",
    title = "{Constraints on leptonically annihilating Dark Matter from reionization and extragalactic gamma background}",
    eprint = "0906.4550",
    archivePrefix = "arXiv",
    primaryClass = "astro-ph.CO",
    doi = "10.1051/0004-6361/200912760",
    journal = "Astron. Astrophys.",
    volume = "505",
    pages = "999--1005",
    year = "2009"
}

@article{Poulin:2015pna,
    author = "Poulin, Vivian and Serpico, Pasquale D. and Lesgourgues, Julien",
    title = "{Dark Matter annihilations in halos and high-redshift sources of reionization of the universe}",
    eprint = "1508.01370",
    archivePrefix = "arXiv",
    primaryClass = "astro-ph.CO",
    doi = "10.1088/1475-7516/2015/12/041",
    journal = "JCAP",
    volume = "12",
    pages = "041",
    year = "2015"
}

@article{Diamanti:2013bia,
    author = "Diamanti, Roberta and Lopez-Honorez, Laura and Mena, Olga and Palomares-Ruiz, Sergio and Vincent, Aaron C.",
    title = "{Constraining Dark Matter Late-Time Energy Injection: Decays and P-Wave Annihilations}",
    eprint = "1308.2578",
    archivePrefix = "arXiv",
    primaryClass = "astro-ph.CO",
    reportNumber = "IFIC-13-54",
    doi = "10.1088/1475-7516/2014/02/017",
    journal = "JCAP",
    volume = "02",
    pages = "017",
    year = "2014"
}

@article{Natarajan:2010dc,
    author = "Natarajan, Aravind and Schwarz, Dominik J.",
    title = "{Distinguishing standard reionization from dark matter models}",
    eprint = "1002.4405",
    archivePrefix = "arXiv",
    primaryClass = "astro-ph.CO",
    doi = "10.1103/PhysRevD.81.123510",
    journal = "Phys. Rev. D",
    volume = "81",
    pages = "123510",
    year = "2010"
}

@article{Valdes:2012zv,
    author = "Valdes, Marcos and Evoli, Carmelo and Mesinger, Andrei and Ferrara, Andrea and Yoshida, Naoki",
    title = "{The nature of dark matter from the global high redshift HI 21 cm signal}",
    eprint = "1209.2120",
    archivePrefix = "arXiv",
    primaryClass = "astro-ph.CO",
    doi = "10.1093/mnras/sts458",
    journal = "Mon. Not. Roy. Astron. Soc.",
    volume = "429",
    pages = "1705--1716",
    year = "2013"
}

@article{Cang:2021owu,
    author = "Cang, Junsong and Gao, Yu and Ma, Yin-Zhe",
    title = "{21-cm constraints on spinning primordial black holes}",
    eprint = "2108.13256",
    archivePrefix = "arXiv",
    primaryClass = "astro-ph.CO",
    doi = "10.1088/1475-7516/2022/03/012",
    journal = "JCAP",
    volume = "03",
    number = "03",
    pages = "012",
    year = "2022"
}

@article{Slatyer:2012yq,
    author = "Slatyer, Tracy R.",
    title = "{Energy Injection And Absorption In The Cosmic Dark Ages}",
    eprint = "1211.0283",
    archivePrefix = "arXiv",
    primaryClass = "astro-ph.CO",
    doi = "10.1103/PhysRevD.87.123513",
    journal = "Phys. Rev. D",
    volume = "87",
    number = "12",
    pages = "123513",
    year = "2013"
}

@article{Slatyer:2015jla,
    author = "Slatyer, Tracy R.",
    title = "{Indirect dark matter signatures in the cosmic dark ages. I. Generalizing the bound on s-wave dark matter annihilation from Planck results}",
    eprint = "1506.03811",
    archivePrefix = "arXiv",
    primaryClass = "hep-ph",
    reportNumber = "MIT-CTP-4682",
    doi = "10.1103/PhysRevD.93.023527",
    journal = "Phys. Rev. D",
    volume = "93",
    number = "2",
    pages = "023527",
    year = "2016"
}

@article{Slatyer:2015kla,
    author = "Slatyer, Tracy R.",
    title = "{Indirect Dark Matter Signatures in the Cosmic Dark Ages II. Ionization, Heating and Photon Production from Arbitrary Energy Injections}",
    eprint = "1506.03812",
    archivePrefix = "arXiv",
    primaryClass = "astro-ph.CO",
    reportNumber = "MIT-CTP-4683",
    doi = "10.1103/PhysRevD.93.023521",
    journal = "Phys. Rev. D",
    volume = "93",
    number = "2",
    pages = "023521",
    year = "2016"
}

@article{Liu:2019bbm,
    author = "Liu, Hongwan and Ridgway, Gregory W. and Slatyer, Tracy R.",
    title = "{Code package for calculating modified cosmic ionization and thermal histories with dark matter and other exotic energy injections}",
    eprint = "1904.09296",
    archivePrefix = "arXiv",
    primaryClass = "astro-ph.CO",
    doi = "10.1103/PhysRevD.101.023530",
    journal = "Phys. Rev. D",
    volume = "101",
    number = "2",
    pages = "023530",
    year = "2020"
}

@article{Evoli:2012zz,
    author = "Evoli, C. and Valdes, M. and Ferrara, A. and Yoshida, N.",
    title = "{Energy deposition by weakly interacting massive particles: a comprehensiv e study}",
    doi = "10.1111/j.1365-2966.2012.20624.x",
    journal = "Mon. Not. Roy. Astron. Soc.",
    volume = "422",
    pages = "420--433",
    year = "2012"
}

@article{Ali-Haimoud:2010hou,
    author = "Ali-Haimoud, Yacine and Hirata, Christopher M.",
    title = "{HyRec: A fast and highly accurate primordial hydrogen and helium recombination code}",
    eprint = "1011.3758",
    archivePrefix = "arXiv",
    primaryClass = "astro-ph.CO",
    doi = "10.1103/PhysRevD.83.043513",
    journal = "Phys. Rev. D",
    volume = "83",
    pages = "043513",
    year = "2011"
}

@article{Lee:2020obi,
    author = {Lee, Nanoom and Ali-Ha\"\i{}moud, Yacine},
    title = "{HYREC-2: a highly accurate sub-millisecond recombination code}",
    eprint = "2007.14114",
    archivePrefix = "arXiv",
    primaryClass = "astro-ph.CO",
    doi = "10.1103/PhysRevD.102.083517",
    journal = "Phys. Rev. D",
    volume = "102",
    number = "8",
    pages = "083517",
    year = "2020"
}

@article{Slatyer:2009yq,
    author = "Slatyer, Tracy R. and Padmanabhan, Nikhil and Finkbeiner, Douglas P.",
    title = "{CMB Constraints on WIMP Annihilation: Energy Absorption During the Recombination Epoch}",
    eprint = "0906.1197",
    archivePrefix = "arXiv",
    primaryClass = "astro-ph.CO",
    doi = "10.1103/PhysRevD.80.043526",
    journal = "Phys. Rev. D",
    volume = "80",
    pages = "043526",
    year = "2009"
}

@article{2018JOSS....3..850M,
       author = {{Murray}, Steven G.},
        title = "{powerbox: A Python package for creating structured fields with isotropic power spectra}",
      journal = {The Journal of Open Source Software},
         year = 2018,
        month = aug,
       volume = {3},
       number = {28},
          eid = {850},
        pages = {850},
          doi = {10.21105/joss.00850},
archivePrefix = {arXiv},
       eprint = {1809.05030},
 primaryClass = {astro-ph.IM},
       adsurl = {https://ui.adsabs.harvard.edu/abs/2018JOSS....3..850M}
}

@article{Sitwell:2013fpa,
    author = "Sitwell, Michael and Mesinger, Andrei and Ma, Yin-Zhe and Sigurdson, Kris",
    title = "{The Imprint of Warm Dark Matter on the Cosmological 21-cm Signal}",
    eprint = "1310.0029",
    archivePrefix = "arXiv",
    primaryClass = "astro-ph.CO",
    doi = "10.1093/mnras/stt2392",
    journal = "Mon. Not. Roy. Astron. Soc.",
    volume = "438",
    number = "3",
    pages = "2664--2671",
    year = "2014"
}

@article{Hooper:2018kfv,
  author = "Hooper, Dan",
  title = "{TASI Lectures on Indirect Searches For Dark Matter}",
  doi = "10.22323/1.333.0010",
  journal = "PoS",
  year = 2019,
  volume = "TASI2018",
  pages = "010"
}

@article{MAGIC:2016xys,
    author = "Ahnen, M. L. and others",
    collaboration = "MAGIC, Fermi-LAT",
    title = "{Limits to Dark Matter Annihilation Cross-Section from a Combined Analysis of MAGIC and Fermi-LAT Observations of Dwarf Satellite Galaxies}",
    eprint = "1601.06590",
    archivePrefix = "arXiv",
    primaryClass = "astro-ph.HE",
    reportNumber = "FERMILAB-PUB-16-283-AE",
    doi = "10.1088/1475-7516/2016/02/039",
    journal = "JCAP",
    volume = "02",
    pages = "039",
    year = "2016"
}

@article{Bergstrom:2013jra,
    author = "Bergstrom, Lars and Bringmann, Torsten and Cholis, Ilias and Hooper, Dan and Weniger, Christoph",
    title = "{New Limits on Dark Matter Annihilation from AMS Cosmic Ray Positron Data}",
    eprint = "1306.3983",
    archivePrefix = "arXiv",
    primaryClass = "astro-ph.HE",
    reportNumber = "FERMILAB-PUB-13-202-A",
    doi = "10.1103/PhysRevLett.111.171101",
    journal = "Phys. Rev. Lett.",
    volume = "111",
    pages = "171101",
    year = "2013"
}

@article{Giesen:2015ufa,
    author = {Giesen, Ga\"elle and Boudaud, Mathieu and G\'enolini, Yoann and Poulin, Vivian and Cirelli, Marco and Salati, Pierre and Serpico, Pasquale D.},
    title = "{AMS-02 antiprotons, at last! Secondary astrophysical component and immediate implications for Dark Matter}",
    eprint = "1504.04276",
    archivePrefix = "arXiv",
    primaryClass = "astro-ph.HE",
    reportNumber = "SACLAY-T13-137, LAPTH-019-15, LAPP-EXP-2015-01",
    doi = "10.1088/1475-7516/2015/9/023",
    journal = "JCAP",
    volume = "09",
    pages = "023",
    year = "2015"
}

@article{Padmanabhan:2005es,
    author = "Padmanabhan, Nikhil and Finkbeiner, Douglas P.",
    title = "{Detecting dark matter annihilation with CMB polarization: Signatures and experimental prospects}",
    eprint = "astro-ph/0503486",
    archivePrefix = "arXiv",
    doi = "10.1103/PhysRevD.72.023508",
    journal = "Phys. Rev. D",
    volume = "72",
    pages = "023508",
    year = "2005"
}

@article{Slatyer:2021qgc,
    author = "Slatyer, Tracy R.",
    title = "{Les Houches Lectures on Indirect Detection of Dark Matter}",
    eprint = "2109.02696",
    archivePrefix = "arXiv",
    primaryClass = "hep-ph",
    reportNumber = "MIT-CTP/5322",
    doi = "10.21468/SciPostPhysLectNotes.53",
    journal = "SciPost Phys. Lect. Notes",
    volume = "53",
    pages = "1",
    year = "2022"
}

@article{Sun:2023acy,
    author = "Sun, Yitian and Foster, Joshua W. and Liu, Hongwan and Mu\~noz, Julian B. and Slatyer, Tracy R.",
    title = "{Inhomogeneous Energy Injection in the 21-cm Power Spectrum: Sensitivity to Dark Matter Decay}",
    eprint = "2312.11608",
    archivePrefix = "arXiv",
    primaryClass = "hep-ph",
    reportNumber = "MIT-CTP/5657, FERMILAB-PUB-23-0816-T-V",
    month = "12",
    year = "2023"
}

@article{Mesinger:2013nua,
    author = "Mesinger, Andrei and Ewall-Wice, Aaron and Hewitt, Jacqueline",
    title = "{Reionization and beyond: detecting the peaks of the cosmological 21 cm signal}",
    eprint = "1310.0465",
    archivePrefix = "arXiv",
    primaryClass = "astro-ph.CO",
    doi = "10.1093/mnras/stu125",
    journal = "Mon. Not. Roy. Astron. Soc.",
    volume = "439",
    number = "4",
    pages = "3262--3274",
    year = "2014"
}

@article{Lidz:2007az,
    author = "Lidz, Adam and Zahn, Oliver and McQuinn, Matthew and Zaldarriaga, Matias and Hernquist, Lars",
    title = "{Detecting the Rise and Fall of 21 cm Fluctuations with the Murchison Widefield Array}",
    eprint = "0711.4373",
    archivePrefix = "arXiv",
    primaryClass = "astro-ph",
    doi = "10.1086/587618",
    journal = "Astrophys. J.",
    volume = "680",
    pages = "962--974",
    year = "2008"
}

@article{Dillon:2013rfa,
    author = "Dillon, Joshua S. and others",
    title = "{Overcoming real-world obstacles in 21 cm power spectrum estimation: A method demonstration and results from early Murchison Widefield Array data}",
    eprint = "1304.4229",
    archivePrefix = "arXiv",
    primaryClass = "astro-ph.CO",
    doi = "10.1103/PhysRevD.89.023002",
    journal = "Phys. Rev. D",
    volume = "89",
    number = "2",
    pages = "023002",
    year = "2014"
}

@article{Pober:2013ig,
    author = "Pober, Jonathan C. and others",
    title = "{Opening the 21cm EoR Window: Measurements of Foreground Isolation with PAPER}",
    eprint = "1301.7099",
    archivePrefix = "arXiv",
    primaryClass = "astro-ph.CO",
    doi = "10.1088/2041-8205/768/2/L36",
    journal = "Astrophys. J. Lett.",
    volume = "768",
    pages = "L36",
    year = "2013"
}

@article{Slatyer:2016qyl,
    author = "Slatyer, Tracy R. and Wu, Chih-Liang",
    title = "{General Constraints on Dark Matter Decay from the Cosmic Microwave Background}",
    eprint = "1610.06933",
    archivePrefix = "arXiv",
    primaryClass = "astro-ph.CO",
    reportNumber = "MIT-CTP-4842",
    doi = "10.1103/PhysRevD.95.023010",
    journal = "Phys. Rev. D",
    volume = "95",
    number = "2",
    pages = "023010",
    year = "2017"
}

@article{Seager:1999bc,
    author = "Seager, Sara and Sasselov, Dimitar D. and Scott, Douglas",
    title = "{A new calculation of the recombination epoch}",
    eprint = "astro-ph/9909275",
    archivePrefix = "arXiv",
    doi = "10.1086/312250",
    journal = "Astrophys. J. Lett.",
    volume = "523",
    pages = "L1--L5",
    year = "1999"
}

@book{dodelson2020modern,
  title={Modern cosmology},
  author={Dodelson, Scott and Schmidt, Fabian},
  year={2020},
  publisher={Academic press}
}

@article{Sokasian:2001xh,
    author = "Sokasian, Aaron and Abel, Tom and Hernquist, Lars. E.",
    title = "{The epoch of helium reionization}",
    eprint = "astro-ph/0112297",
    archivePrefix = "arXiv",
    doi = "10.1046/j.1365-8711.2002.05291.x",
    journal = "Mon. Not. Roy. Astron. Soc.",
    volume = "332",
    pages = "601",
    year = "2002"
}

@article{Facchinetti:2023slb,
    author = "Facchinetti, Ga\'etan and Lopez-Honorez, Laura and Qin, Yuxiang and Mesinger, Andrei",
    title = "{21cm signal sensitivity to dark matter decay}",
    eprint = "2308.16656",
    archivePrefix = "arXiv",
    primaryClass = "astro-ph.CO",
    reportNumber = "ULB-TH/23-09",
    doi = "10.1088/1475-7516/2024/01/005",
    journal = "JCAP",
    volume = "01",
    pages = "005",
    year = "2024"
}

@article{Bringmann:2009vf,
    author = "Bringmann, Torsten",
    title = "{Particle Models and the Small-Scale Structure of Dark Matter}",
    eprint = "0903.0189",
    archivePrefix = "arXiv",
    primaryClass = "astro-ph.CO",
    doi = "10.1088/1367-2630/11/10/105027",
    journal = "New J. Phys.",
    volume = "11",
    pages = "105027",
    year = "2009"
}

@article{Profumo:2006bv,
    author = "Profumo, Stefano and Sigurdson, Kris and Kamionkowski, Marc",
    title = "{What mass are the smallest protohalos?}",
    eprint = "astro-ph/0603373",
    archivePrefix = "arXiv",
    doi = "10.1103/PhysRevLett.97.031301",
    journal = "Phys. Rev. Lett.",
    volume = "97",
    pages = "031301",
    year = "2006"
}

@article{vandenAarssen:2012ag,
    author = "van den Aarssen, Laura G. and Bringmann, Torsten and Goedecke, Yasar C",
    title = "{Thermal decoupling and the smallest subhalo mass in dark matter models with Sommerfeld-enhanced annihilation rates}",
    eprint = "1202.5456",
    archivePrefix = "arXiv",
    primaryClass = "hep-ph",
    doi = "10.1103/PhysRevD.85.123512",
    journal = "Phys. Rev. D",
    volume = "85",
    pages = "123512",
    year = "2012"
}

@article{Hofmann:2001bi,
    author = "Hofmann, Stefan and Schwarz, Dominik J. and Stoecker, Horst",
    title = "{Damping scales of neutralino cold dark matter}",
    eprint = "astro-ph/0104173",
    archivePrefix = "arXiv",
    doi = "10.1103/PhysRevD.64.083507",
    journal = "Phys. Rev. D",
    volume = "64",
    pages = "083507",
    year = "2001"
}

@article{Sheth:2001dp,
    author = "Sheth, Ravi K. and Tormen, Giuseppe",
    title = "{An Excursion Set Model of Hierarchical Clustering : Ellipsoidal Collapse and the Moving Barrier}",
    eprint = "astro-ph/0105113",
    archivePrefix = "arXiv",
    reportNumber = "FERMILAB-PUB-01-061-A",
    doi = "10.1046/j.1365-8711.2002.04950.x",
    journal = "Mon. Not. Roy. Astron. Soc.",
    volume = "329",
    pages = "61",
    year = "2002"
}

@article{Sheth:1999mn,
    author = "Sheth, Ravi K. and Tormen, Giuseppe",
    title = "{Large scale bias and the peak background split}",
    eprint = "astro-ph/9901122",
    archivePrefix = "arXiv",
    doi = "10.1046/j.1365-8711.1999.02692.x",
    journal = "Mon. Not. Roy. Astron. Soc.",
    volume = "308",
    pages = "119",
    year = "1999"
}

@article{efstathiou1992cobe,
  title={COBE background radiation anisotropies and large-scale structure in the universe},
  author={Efstathiou, G and Bond, JR and White, Simon DM},
  journal={Monthly Notices of the Royal Astronomical Society},
  volume={258},
  number={1},
  pages={1P--6P},
  year={1992},
  publisher={Oxford University Press Oxford, UK}
}

@article{Sanchez-Conde:2013yxa,
    author = "S\'anchez-Conde, Miguel A. and Prada, Francisco",
    title = "{The flattening of the concentration\textendash{}mass relation towards low halo masses and its implications for the annihilation signal boost}",
    eprint = "1312.1729",
    archivePrefix = "arXiv",
    primaryClass = "astro-ph.CO",
    doi = "10.1093/mnras/stu1014",
    journal = "Mon. Not. Roy. Astron. Soc.",
    volume = "442",
    number = "3",
    pages = "2271--2277",
    year = "2014"
}

@article{Maccio:2006wpz,
    author = "Maccio', Andrea V. and Dutton, Aaron A. and van den Bosch, Frank C. and Moore, Ben and Potter, Doug and Stadel, Joachim",
    title = "{Concentration, Spin and Shape of Dark Matter Haloes: Scatter and the Dependence on Mass and Environment}",
    eprint = "astro-ph/0608157",
    archivePrefix = "arXiv",
    doi = "10.1111/j.1365-2966.2007.11720.x",
    journal = "Mon. Not. Roy. Astron. Soc.",
    volume = "378",
    pages = "55--71",
    year = "2007"
}

@article{Maccio:2008pcd,
    author = "Maccio', Andrea V. and Dutton, Aaron A. and Bosch, Frank C. van den",
    title = "{Concentration, Spin and Shape of Dark Matter Haloes as a Function of the Cosmological Model: WMAP1, WMAP3 and WMAP5 results}",
    eprint = "0805.1926",
    archivePrefix = "arXiv",
    primaryClass = "astro-ph",
    doi = "10.1111/j.1365-2966.2008.14029.x",
    journal = "Mon. Not. Roy. Astron. Soc.",
    volume = "391",
    pages = "1940--1954",
    year = "2008"
}

@article{Sarkar:2022dvl,
    author = "Sarkar, Debanjan and Flitter, Jordan and Kovetz, Ely D.",
    title = "{Exploring delaying and heating effects on the 21-cm signature of fuzzy dark matter}",
    eprint = "2201.03355",
    archivePrefix = "arXiv",
    primaryClass = "astro-ph.CO",
    doi = "10.1103/PhysRevD.105.103529",
    journal = "Phys. Rev. D",
    volume = "105",
    number = "10",
    pages = "103529",
    year = "2022"
}

@article{Koopmans:2015sua,
    author = "Koopmans, L. V. E. and others",
    editor = "Bourke, Tyler L. and others",
    title = "{The Cosmic Dawn and Epoch of Reionization with the Square Kilometre Array}",
    eprint = "1505.07568",
    archivePrefix = "arXiv",
    primaryClass = "astro-ph.CO",
    doi = "10.22323/1.215.0001",
    journal = "PoS",
    volume = "AASKA14",
    pages = "001",
    year = "2015"
}

@article{Lopez-Honorez:2013cua,
    author = "Lopez-Honorez, Laura and Mena, Olga and Palomares-Ruiz, Sergio and Vincent, Aaron C.",
    title = "{Constraints on dark matter annihilation from CMB observationsbefore Planck}",
    eprint = "1303.5094",
    archivePrefix = "arXiv",
    primaryClass = "astro-ph.CO",
    reportNumber = "IFIC-13-016, CFTP-13-007",
    doi = "10.1088/1475-7516/2013/07/046",
    journal = "JCAP",
    volume = "07",
    pages = "046",
    year = "2013"
}
\end{document}